\documentclass[a4paper,scale=0.8,onecolumn,nobibnotes,nofootinbib]{revtex4}
\usepackage{enumitem}
\usepackage{amsmath,amssymb}
\usepackage{graphicx} 
\usepackage{epstopdf}
\usepackage{hyperref}
\usepackage{geometry}
\usepackage{subfigure}

\hypersetup{
    colorlinks=true,
    citecolor=blue,
    linkcolor=red,
    filecolor=magenta,
    urlcolor=cyan,}

\newcommand{\be}{\begin{equation}}
\newcommand{\ee}{\end{equation}}

\begin{document}

\title{Joule-Thomson expansion of charged dilatonic black holes}
\author{Meng-Yao Zhang\footnote{gs.myzhang21@gzu.edu.cn}$^{1}$, Hao Chen\footnote{ haochen1249@yeah.net}$^{2}$, Hassan Hassanabadi\footnote{h.hasanabadi@shahroodut.ac.ir}$^{3}$,  Zheng-Wen Long\footnote{zwlong@gzu.edu.cn (Corresponding author)}$^{4}$, and Hui Yang\footnote{huiyang@gzu.edu.cn (Corresponding author)}$^{1}$}
\affiliation{$^1$ School of Mathematics and Statistics, Guizhou University, Guiyang, 550025, China.\\
$^{2}$ School of Physics and Electronic Science, Zunyi Normal University, Zunyi 563006, China.\\
$^{3}$ Faculty of Physics, Shahrood University of Technology, Shahrood, Iran.\\
$^{4}$ College of Physics, Guizhou University, Guiyang, 550025, China.
}

\date{\today}

\begin{abstract}
\textbf{Abstract:} Based on the Einstein-Maxwell theory, the Joule-Thomson (J-T) expansion of charged dilatonic black holes (the solutions are neither flat nor AdS) in $(n+1)$-dimensional spacetime is studied herein. To this end, we analyze the effects of the dimension $n$ and dilaton field $\alpha$ on J-T expansion. An explicit expression for the J-T coefficient is derived, and consequently, a negative heat capacity is found to lead to a cooling process. In contrast to its effect on the dimension, the inversion curve decreases with charge $Q$ at low pressures, whereas the opposite effect is observed at high pressures. We can observe that with an increase in the dimension $n$ or parameter $\alpha$, both the pressure cut-off point and the minimum inversion temperature $T_{min}$ change. Moreover, we analyze the ratio $T_{min}/T_{c}$ numerically and discover that the ratio is independent of charge; however, it depends on the dilaton field and dimension: for $n=3$ and $\alpha=0$, the ratio is 1/2. The dilaton field is found to enhance the ratio. In addition, we identify the cooling-heating regions by investigating the inversion and isenthalpic curves, and the behavior of the minimum inversion mass $M_{min}$ indicates that this cooling-heating transition may not occur under certain special conditions.
\end{abstract}
\maketitle

\section{Introduction}
Einstein's theory of general relativity is among the two most important breakthroughs of physics in this century, and the theory creatively identified gravity as an effect of the curvature of space-time caused by the presence of matter.  The cosmological constant in the Einstein field equation mainly describes the state of the universe. The general theory of relativity has facilitated profound understanding of the nature and interrelation of time, space, matter and gravity, and it has been widely acknowledged. The development of general relativity relies heavily on the solutions of Einstein's gravitational field equations and their physical interpretations.  Therefore, the solutions of the gravitational field equations and the analysis of their properties are important components of the theory of gravity, and every new gravitational solution plays an important role in our understanding of the nature of space-time and gravity.  In addition, the study of high-dimensional gravitational solutions also contributes toward our understanding of the spatio-temporal stability and dimensional dependence of gravitational properties.

A black hole (BH) is a direct result of general relativity; a notable sign of a BH exists is the event horizon. The event horizon distinguishes between the inside and the outside of a BH, and particles that enter the event horizon can no longer escape. In the past few decades, pioneering studies \cite{P1,P2,P3} have established a deep and fundamental link between black holes and general thermodynamic systems.  In general relativity, black holes can be realized by solving Einstein's field equations, and as thermodynamic systems with a physical temperature and entropy,  black holes have a huge impact on the understanding of quantum gravity \cite{Q1,Q2,Q3,Q4}.  Following decades of development, the thermodynamic properties of black holes have attracted extensive research interest; these include quantum effects on the thermodynamic properties of black holes \cite{ch1,ch2,ch3,ch4}, entropy \cite{E1,E2}, thermodynamic phase transition \cite{PT1,PT2,PT3}, thermal and non-thermal radiation \cite{R1,R2,R3}, and various criticalities \cite{C1,C2,C3,C4,C5,C6,C7}.
In the extended phase space, the negative cosmological constant is treated as a variable thermodynamic pressure of a BH system, its conjugate quantity is naturally regarded as the BH volume in the mechanics of AdS black holes, and the mass of the BH is interpreted as the enthalpy of spacetime and not as the internal energy \cite{A1}. The above ideas have been subsequently generalized and applied to black holes in the Lovelock theory. For example, in \cite{A2}, the authors studied AdS black holes based on the Lovelock theory, and they used geometric methods to obtain the Smarr formula. They also proved that the variations of the Lovelock couplings were the extension of the first law. In \cite{A3}, an explicit formula for the ADM mass and free energy of AdS black holes was obtained based on the Lovelock theory. In addition, a study on the dilaton-AdS black holes revealed that the mass of a BH represents its enthalpy \cite{A4}. These innovative studies have led to remarkable consequences in the thermodynamics research of AdS black holes in the extended phase space.
The thermodynamic pressure in the following units $(G_{N} = \hbar  = c = \kappa = 1)$ can be expressed by the cosmological constant as follows:
\begin{equation}
\begin{aligned}
P=-\frac{\Lambda }{8\pi } \nonumber.
\end{aligned}
\end{equation}
Moreover, the associated volume is
\begin{equation}
\begin{aligned}
V=\frac{\partial M}{\partial P}  \nonumber.
\end{aligned}
\end{equation}
 Accordingly, related thermodynamic properties of some classical systems can be obtained based on analogy in the BH model \cite{M1,M2} and the behavior of BH like van der Waals fluids \cite{F1,F2,F3}. The similarities between black holes in the AdS space indicate a significant correspondence between AdS and CFT \cite{ADS}. Indeed, from the perspective of AdS/CFT, in \cite{Q3}, the authors demonstrated that the asymptotic AdS space-time BH is perfectly consistent with the canonical dual description of the double thermal field theory, and this physical phenomenon is known as the Hawking-Page phase transition. More interestingly, other theories, such as those of rotating and hairy black holes \cite{BH1,BH2} and the m-theory \cite{MT1,MT2}, have been widely generalized. To further reinforce the resemblance to van der Waals fluids, researchers have obtained additional exotic results, such as the behavior of the quasi-normal modes \cite{QU1,QU2}, holographic heat engines \cite{HE1}, and chaotic structures \cite{ST1}.

In classical thermodynamics, it is well known that the process of gas movement from a region of high pressure to a region of low pressure with an equal velocity is known as J-T expansion. Based on this theory,  \"Okc\"u and E. Aydiner  were the first to study the J-T effect of charged AdS black holes \cite{J1}. Following this, J-T expansion emerged as an active field and received considerable attention, and it was subsequently extended to the study of all types of black holes \cite{J2,J3,J4,J5,J6,J7,J8}. These comprehensive studies have revealed that the heating and cooling regions can be perfectly identified based on the T-P diagram. The intersection of the inversion and the enthalpic curves can distinguish the heating-cooling regions of the black hole. When the J-T coefficient $\mu>0$, the cooling region is obtained, and when $\mu<0$, the heating region is obtained. Moreover, the ratio between $T_{min}$ and the critical temperature $T_{c}$ is an important aspect of J-T expansion, and this ratio is 1/2 for an RN-AdS BH \cite{J1}. For other typical black holes, this ratio has been evaluated in previous studies \cite{J2,J3,J5}. In this study,  the ratio was restored to 1/2 in some special cases.

The scalar-tensor theory of gravity has emerged as a research focus. The minimal coupling between the scalar fields and gravity was first considered in \cite{RE1}. As one of the most interesting scalar fields, dilaton gravity is regarded as an effective field theory description of the string theory at low energy \cite{RE2}. This theory provides an approach to correct Einstein's theory of gravity. The coupling of the dilaton field with the Einstein-Maxwell theory affects the causality of the space-time structure \cite{RE3}. In recent years, inspired by the idea of AdS/CFT duality, the research on AdS dilaton black holes has attracted increasing attention. In addition to using the cosmological constant as the potential energy, researchers have been particularly interested in studying the Liouville-type dilaton potential energy, which originates from supersymmetry breaking in the high-dimensional supergravity model. Consequently, studies have revealed that for only one or two Liouville type potentials, neither asymptotically flat nor (A)dS BH solutions can be obtained \cite{RE4}. For example, the charged dilatonic BH solutions in $(n+1)$-dimensional spacetime \cite{RE5} and many other interesting physical phenomena of such black holes have been investigated; these include thermal stability \cite{RE5}, various phase transitions \cite{RE6,RE7,RE8} and criticality \cite{RE9,RE10}. Despite this, we believe that the thermodynamic properties of this typical class of dilaton black holes remain to be explored. Hence, it is very natural to explore the effect of the dilaton parameter on the J-T expansion of charged black holes based on the Einstein-Maxwell-dilaton gravity theory. Additional discussions on the effect of the scalar field on J-T expansion can be found elsewhere(see \cite{J6,RE12,RE13} for more details). We believe that our results can shed some light on the possible extensions of AdS/CFT correspondence.

The aim of this study was to analyze the J-T expansion of charged dilatonic black holes and examine the effects of BH characteristics. The remainder of this manuscript is structured as follows. In section II, we provide a brief review of $(n+1)$-dimensional charged dilatonic black holes in the presence of Einstein-dilaton gravity. In section III, we discuss four important aspects of J-T expansion. First, we derive the J-T coefficient. Following this, we obtain equations for the inversion pressure $P_{i}$, minimum radius $r_{min}$ and minimum inversion temperature $T_{min}$. We also numerically analyze the ratio between $T_{min}$ and the critical temperature $T_{c}$ for charged dilatonic black holes. Furthermore, isenthalpic curves are investigated. Finally, some concluding remarks are presented in the final section.
\section{brief introduction to charged dilatonic black holes} \label{sec2}
In this section, we begin with a brief introduction to the action of $(n + 1)$-dimensional$(n\ge3)$ spacetime with a scalar dilaton field in the presence of the Einstein-dilaton gravity. From \cite{RE4}, we have the following:
\begin{equation}
\begin{aligned}
\mathcal{I}= \frac{1}{16 \pi} \int d^{n+1} x \sqrt{-g}\left(\mathcal{R}-\frac{4}{n-1}(\nabla \varphi)^{2}-\mathcal{V}(\varphi)\right.\left.-e^{-4 \alpha \varphi /(n-1)} F_{\mu \nu} F^{\mu \nu}\right),
\end{aligned}
\end{equation}
where $\mathcal{R}$ is the Ricci scalar curvature, $F_{\mu \nu} = \partial_{[\mu A_{ \nu } ]}$, $A_{ \nu }$ indicates the electromagnetic potential, $\alpha$ is the coupling constant of the scalar, $\varphi$ is the dilaton field, and $\mathcal{V}(\varphi)$ is a potential for $\varphi$, which should be chosen with the following form \cite{RE4, RE5}:
\begin{equation}
\mathcal{V}(\varphi)=2\Lambda e^{4\alpha \varphi /(n-1)} +\frac{k(n-1)(n-2)\alpha ^{2} }{b^{2}(\alpha ^{2}-1)  } e^{4 \varphi /(n-1)\alpha},
\end{equation}
where $\Lambda$ is a free parameter playing the role of the cosmological constant, which is considered it as a thermodynamic quantity in this study. For convenience, $\Lambda$ is redefined as $\Lambda=-n(n-1)/2l^{2}$ \cite{RE5}, where $l$ denotes the length scale, and $b$ is a  nonzero positive arbitrary constant. If we set $(\alpha  = 0)$, Eq.(2) reduces to $\mathcal{V}(\varphi)=2\Lambda$. The corresponding topological BH solution transforms into \cite{RE5}
\begin{equation}
d s^{2}=-f(r) d t^{2}+\frac{d r^{2}}{f(r)}+r^{2} R(r)^{2} d \Omega_{k}^{2},
\end{equation}
where
\begin{equation}
\begin{aligned}
f(r)=& \frac{2 \Lambda\left(\alpha^{2}+1\right)^{2} b^{2 \gamma}}{(n-1)\left(\alpha^{2}-n\right)} r^{2(1-\gamma)}-\frac{k(n-2)\left(\alpha^{2}+1\right)^{2} b^{-2 \gamma} r^{2 \gamma}}{\left(\alpha^{2}-1\right)\left(\alpha^{2}+n-2\right)} \\
&-\frac{m}{r^{(n-1)(1-\gamma)-1}}+\frac{2 q^{2}\left(\alpha^{2}+1\right)^{2} b^{-2(n-2) \gamma}}{(n-1)\left(\alpha^{2}+n-2\right)} r^{2(n-2)(\gamma-1)}.
\end{aligned}
\end{equation}
$\gamma =\alpha ^{2} /(\alpha ^{2}+1 )$; the parameter $m$ is the integration constant associated with mass, and $q$ is the charge parameter of the black holes. Here $k$ implies an $(n-1)$-dimensional hypersurface $d \Omega_{k}^{2}$, where $k$ have the following values: -1, 0, 1.
According to the definition of mass, two related expressions can be given as follows \cite{RE5}
 \begin{equation}
M=\frac{b^{(n-1) \gamma}(n-1) \omega_{n-1}}{16 \pi\left(\alpha^{2}+1\right)} m, \quad Q=\frac{q \omega_{n-1}}{4 \pi}.
\end{equation}
$\omega_{n-1}$ denotes the volume of a unit $(n-1)$-sphere with a unit radius and can be obtained from the following relation:
\begin{equation}
\omega_{n-1}=\frac{2 \pi^{\frac{n}{2}}}{\Gamma\left(\frac{n}{2}\right)}.
\end{equation}
Based on this, by solving for $f(r)|_{r=r_{+}}=0$ in Eq. (4), we can express  $m$ in terms of $r_{+}$ and substitute it in Eq. (5). We can obtain the temperature of the BH as follows:
\begin{align}
T&= \frac{f^{\prime}\left(r_{+}\right)}{4 \pi}=-\frac{k(n-2)\left(\alpha^{2}+1\right) b^{-2 \gamma}}{4 \pi\left(\alpha^{2}-1\right)} r_{+}^{2 \gamma-1}\cr
&-\frac{\Lambda\left(\alpha^{2}+1\right) b^{2 \gamma}}{2 \pi(n-1)} r_{+}^{1-2 \gamma}-\frac{q^{2}\left(\alpha^{2}+1\right) b^{-2(n-2) \gamma}}{2 \pi(n-1)} r_{+}^{(2 n-3)(\gamma-1)-\gamma} .
\end{align}	
By integrating the first law $dS= \frac{1}{T}dM$ and using Eqs. (5-7), the entropy for this BH can be obtained as		
\begin{align}
S=\frac{b^{(n-1)\gamma}{r}^{(n-1)(1-\gamma)}_{+} \omega_{n-1}}{4},
\end{align}
Moreover, the electric potential \cite{RE5} is
\begin{equation}
U=\frac{q b^{(3-n) \gamma}}{\lambda r_{+}^{\lambda}},
\end{equation}
where $\lambda=(n-3)(1-\gamma)+1$. Based on the first law of thermodynamics in the extended phase space,  these thermodynamic quantities satisfy
\begin{equation}
d M=T d S+U d Q+V d P.
\end{equation}
Note that the thermodynamic volume is obtained as
\begin{equation}
V=\int 4 S d r_{+}=\frac{\left(1+\alpha^{2}\right) b^{(n-1)\gamma} \omega_{n-1}}{n+\alpha^{2}} r_{+}^{\left(n+\alpha^{2}\right) /\left(1+\alpha^{2}\right)}.
\end{equation}
The Smarr formula connects the mass of a BH with other geometrical and dynamical parameters of the BH, including the spin, charge, and electromagnetic potential. The formula has the following form \cite{Smarr1,A1}:
\begin{equation}
M=\frac{(n-1)(1-\gamma )}{\lambda }  T S+U Q + \frac{4\gamma -2}{\lambda } V P,
\end{equation}
where $P$ is the thermodynamic pressure given by \cite{RE10}
\begin{equation}
P=-\frac{(n+\alpha ^{2} )b^{2\gamma } }{8\pi (n-\alpha ^{2})r_{+}^{2\gamma }  } \Lambda.
\end{equation}
One may note that $P$ depends on the dilaton parameter. In the absence of a dilaton $(\alpha=0)$, the above pressure reduces to $P=-\frac{\Lambda}{8\pi}$, which corresponds to the  pressure of a Reissner-Nordstrum BH. For charged dilatonic black holes, $P-V$ criticality has been carefully investigated in Ref. \cite{RE10}, and the critical temperature $T_{c}$ is given as
\begin{equation}
\begin{aligned}
T_{c}=& {\left[\frac{(n-1)}{8 \pi^{2-n}Q^{2}\Gamma (\frac{n}{2} )^{2} \left(n-1+\alpha^{2}\right)}\right]^{(1-2 \gamma) / 2 \lambda } \times\left[\frac{\left(2 n-3+\alpha^{2}\right)}{k(n-2)}\right]^{[(2 n-3)(\gamma-1)-\gamma] / 2 \lambda } } \\
& \times\left[\frac{\left(\alpha^{2}+n-2\right) b^{-\gamma(n-1) / \lambda}}{\pi\left(1-\alpha^{2}\right)}\right].
\end{aligned}
\end{equation}
We restrict the dilaton parameter to $\alpha<1$  to ensure a positive value of the critical temperature.
\section{J-T coefficient} \label{sec3}
\begin{figure}[htbp]
	\centering
	\subfigure[$\alpha=0$]{
    \begin{minipage}[t]{0.4\linewidth}
		\centering
		\includegraphics[width=2.6in,height=1.6in]{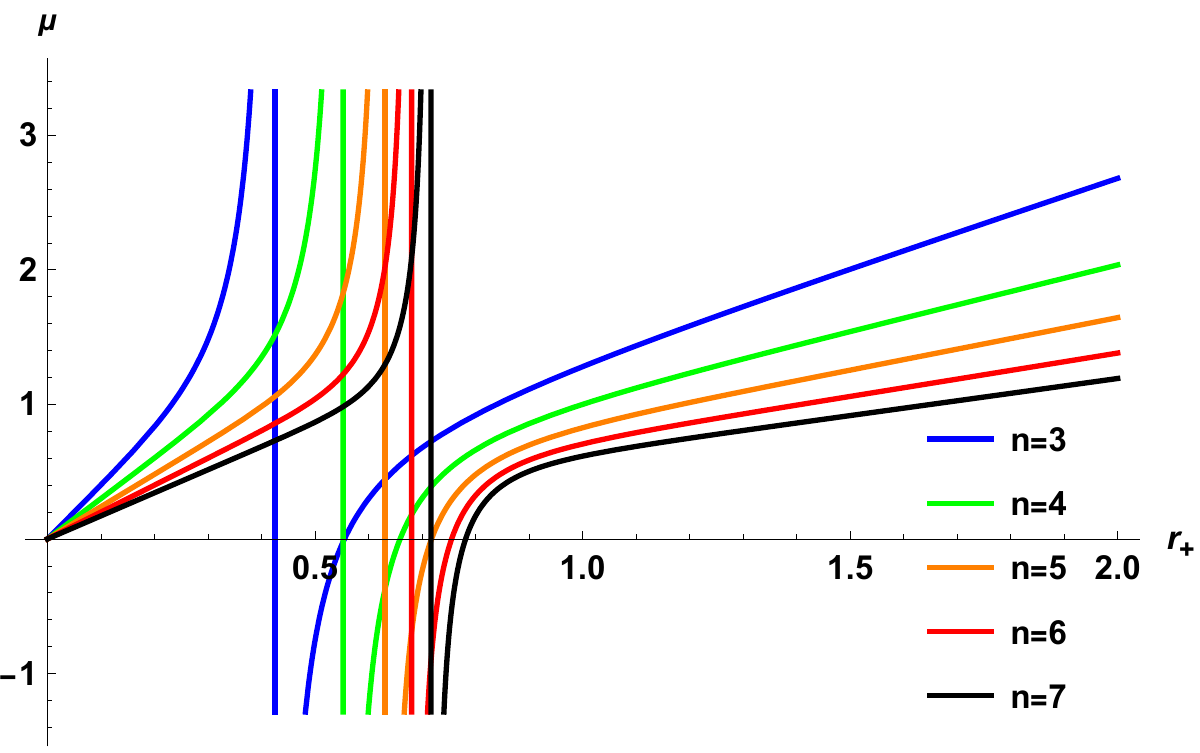}
		\end{minipage}%
            }%
    \subfigure[$\alpha=0$]{
    \begin{minipage}[t]{0.4\linewidth}
		\centering
		\includegraphics[width=2.6in,height=1.6in]{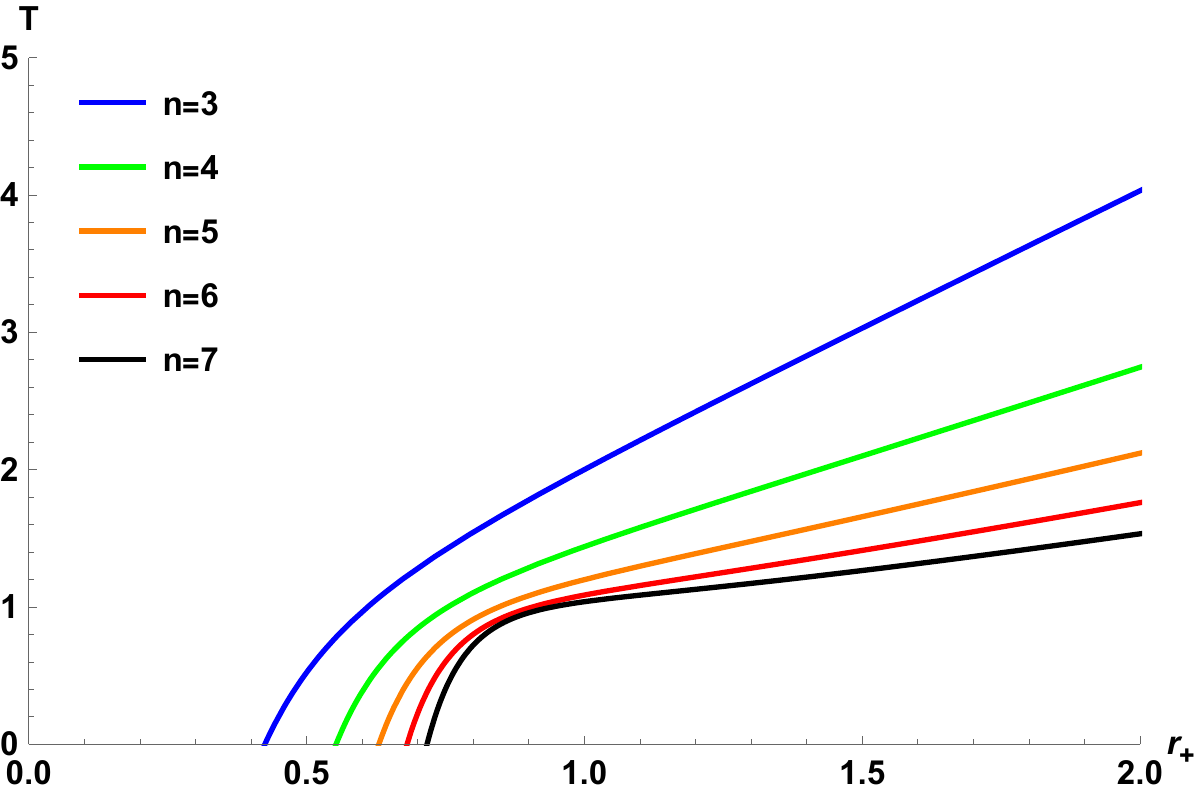}
		\end{minipage}%
            }%

      \subfigure[$\alpha=0.4$]{
    \begin{minipage}[t]{0.4\linewidth}
		\centering
		\includegraphics[width=2.6in,height=1.6in]{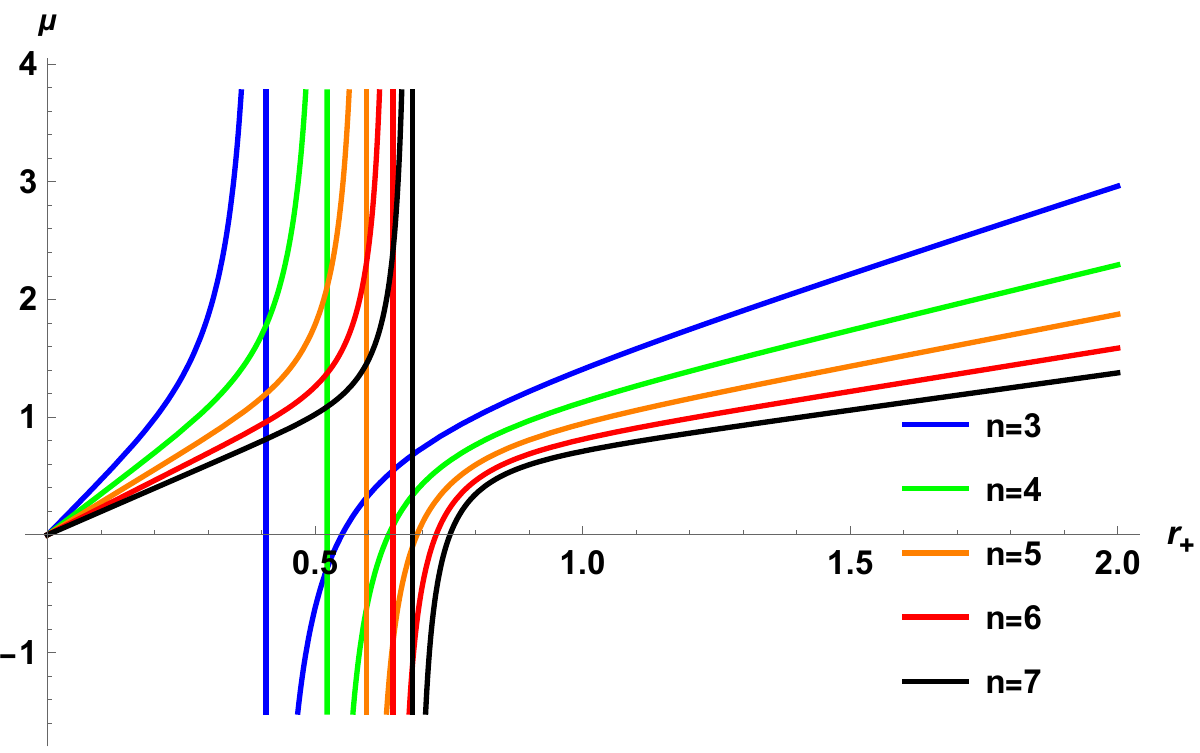}
		\end{minipage}%
            }%
            \subfigure[$\alpha=0.4$]{
    \begin{minipage}[t]{0.4\linewidth}
		\centering
		\includegraphics[width=2.6in,height=1.6in]{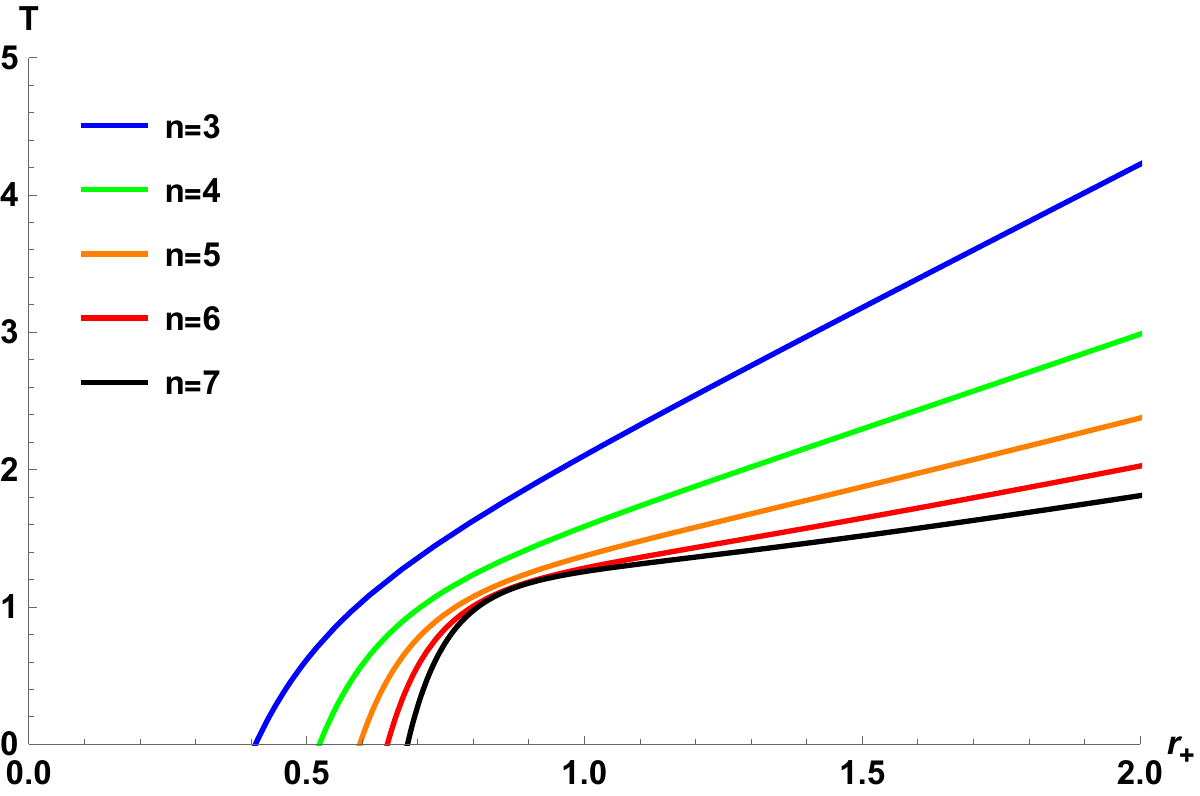}
		\end{minipage}%
            }%

            \subfigure[$\alpha=0.8$]{
    \begin{minipage}[t]{0.4\linewidth}
		\centering
		\includegraphics[width=2.6in,height=1.6in]{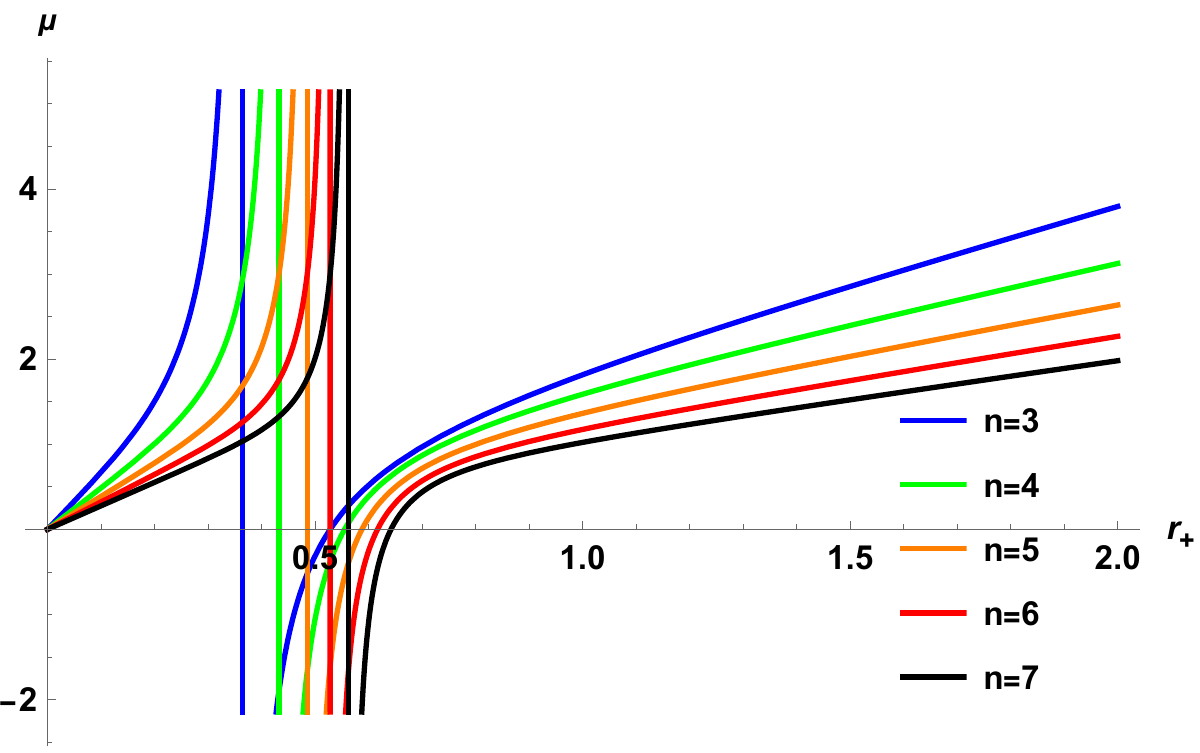}
		\end{minipage}%
            }%
            \subfigure[$\alpha=0.8$]{
    \begin{minipage}[t]{0.4\linewidth}
		\centering
		\includegraphics[width=2.6in,height=1.6in]{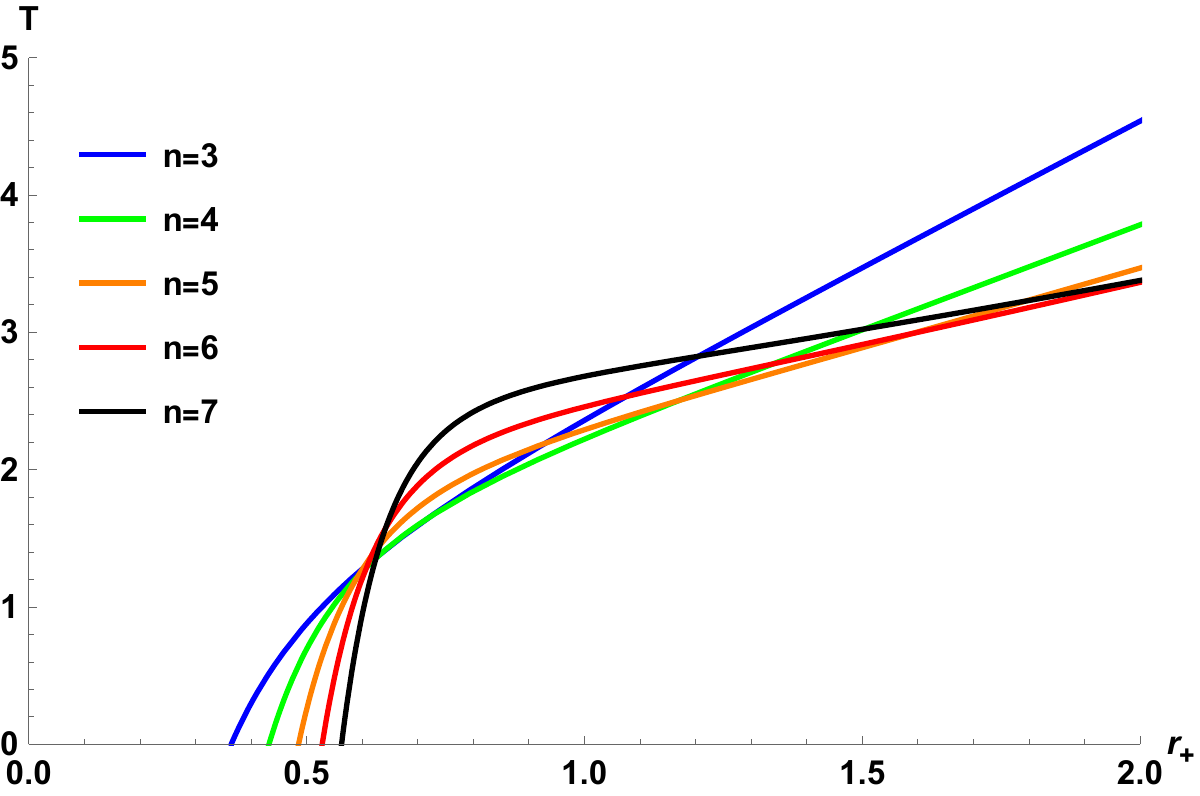}
		\end{minipage}%
            }%
     \centering
     \caption{J-T coefficient $\mu$ and Hawking temperature $T$ of charged dilatonic black holes for $P=1$, $q=1$, $b=1$, and $k=1$. From the top to bottom, the curves correspond to $\alpha=0, 0.4, 0.8$.}
\end{figure}
Next, we investigate the J-T expansion for charged dilatonic black holes. Interestingly, J-T expansion is also known as throttling expansion.  In this process, the actual gas is subjected to a certain pressure and undergoes adiabatic expansion into the low-pressure section of a heat insulation pipe through a porous plug, causing a change in the temperature of the fluid. The enthalpy is maintained constant during the expansion.  The rate of change of temperature with pressure is generally defined as the J-T coefficient. For a fixed charge, the J-T coefficient is given by \cite{J1}
\begin{equation}
\mu=\left(\frac{\partial T}{\partial P}\right)=\frac{1}{C_{P}}\left[T\left(\frac{\partial V}{\partial T}\right)-V\right].
\end{equation}
Here, the property of the J-T coefficient is generally used to indicate whether the gas cools$(\mu>0)$ or warms$(\mu<0)$ after the adiabatic expansion, where $C_{P}$ is the heat capacity expressed as
\begin{equation}
C_{P}=T\left(\frac{\partial S}{\partial T}\right).
\end{equation}
By setting $\varepsilon=\left(n+\alpha^{2}\right) /\left(1+\alpha^{2}\right)$, we can rewrite the thermodynamic volume as
\begin{equation}
V=\frac{ b^{(n-1)}\omega_{n-1}}{\varepsilon} r^{\varepsilon}_{+}.
\end{equation}
Thereafter, substituting Eqs. (7), (8), and (17) into Eqs. (15) and (16) yields
\begin{equation}
\mu =\frac{[ 4r^{1+\varepsilon+n(\gamma -1)-\gamma   }_{+} ( 16\pi P(n-\alpha ^{2} )(\alpha ^{2}-1)-A\omega q^{2}r^{2+2n(\gamma -1)-4\gamma  }_{+} -B(2\gamma-1)r^{2\gamma-2}_{+})(\varepsilon Y(r_{+})-1)   ] }
{[\varepsilon (n-1)(1-\gamma )(16\pi P(n-\alpha ^{2} )(\alpha ^{2}-1)-A q^{2}r^{2+2n(\gamma -1)-4\gamma  }_{+}-Br^{2\gamma-2}_{+})]} ,
\end{equation}
where
\begin{equation}
\omega =3+2n(\gamma -1)-4\gamma ,
\end{equation}
\begin{equation}
A=2b^{-2(n-2)\gamma } (n+\alpha ^{2} )(\alpha ^{2}-1),
\end{equation}
\begin{equation}
B=b^{-2\gamma }(n-2)(n-1)(n+\alpha ^{2} )k,
\end{equation}
\begin{equation}
Y(r_{+}) =\frac{ 16\pi P(n-\alpha ^{2} )(\alpha ^{2}-1)-Aq^{2}r^{2+2n(\gamma -1)-4\gamma  }_{+} -Br^{2\gamma-2}_{+}}
{16\pi P(n-\alpha ^{2} )(\alpha ^{2}-1)-A\omega  q^{2}r^{2+2n(\gamma -1)-4\gamma  }_{+}-B(2\gamma-1)r^{2\gamma-2}_{+})} .
\end{equation}
According to Eq.(7) and Eqs.(18-22), the behaviors of $\mu$ and $T$ with the horizon radius $r_{+}$ are depicted in FIG. 1. From (a) to (f), by setting $P=1$, $q=1$, $b=1$, and $k=1$, we can observe the constraints of the dimension and  dilaton field parameter $\alpha$ on the J-T coefficient $\mu$ and Hawking temperature $T$. Divergence points (vertical curves) and zero points can be noted in each set of the graphs.  The corresponding event horizon radius increases with an increase in the dimension $n$ but decreases with an increase in the parameter $\alpha$.  The $\mu-r_{+}$ scatter coincides with the $T-r_{+}$ zero point, which is interpreted as the reversal point of the cooling and heating processes.  By comparing FIGs. 1(b), 1(d), and 1(f), we observe that as the parameter $\alpha$ increases and approaches 1 (as shown in FIGs.  1(f)), for the positive and non-zero point of the Hawking temperature, with an increase in the dimension $n$, the horizon radius first increases, then decreases, and finally increases again. Therefore, we confidently suggest that the parameter $\alpha$ affects the thermodynamic properties of the system.

Next, we explore the possible influence of the negative heat capacity on J-T expansion. We present the graphical behavior of the J-T coefficient, heat capacity $C_{P}$, and square bracket item $\left[T\left(\frac{\partial V}{\partial T}\right)-V\right]$ in FIG.2; the results indicate the presence of only one root for the heat capacity. For $r_{+}<r_{0}$ (in which $C_{P}(r_{+}=r_{0})=0$), the heat capacity $C_{P}$ and the term $\left[T\left(\frac{\partial V}{\partial T}\right)-V\right]$ are negative, and the J-T coefficient is positive. For $r_{0}<r_{+}<r_{1}$ (in which $\left[T\left(\frac{\partial V}{\partial T}\right)-V\right](r_{+}=r_{1})=0$), the heat capacity $C_{P}$ is positive, the term $\left[T\left(\frac{\partial V}{\partial T}\right)-V\right]$ is negative, and the J-T coefficient is negative. For $r_{+}>r_{1}$, the heat capacity $C_{P}$ and the term $\left[T\left(\frac{\partial V}{\partial T}\right)-V\right]$ are positive, and J-T coefficient is also positive. These graphical behaviors are completely consistent with Eq. (15). Consequently, a negative heat capacity corresponds to a positive J-T coefficient, which indicates the occurrence of cooling process.
\begin{figure}[htbp]
		\centering
		\includegraphics[width=3in,height=2in]{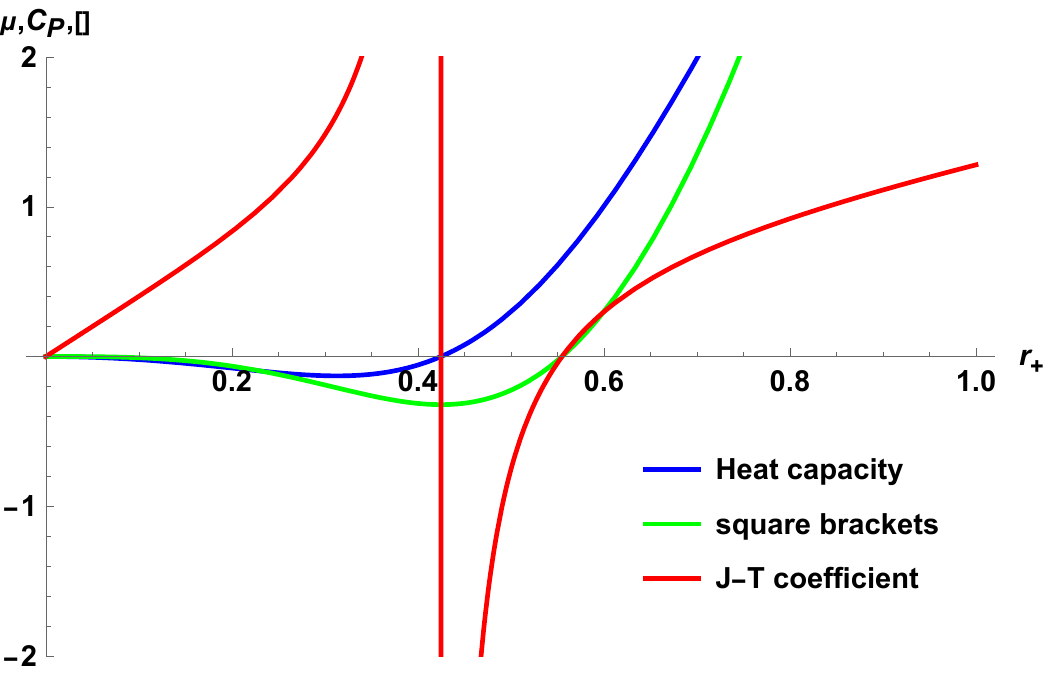}
     \caption{Graphical behavior of the J-T coefficient, heat capacity $C_{P}$, and square bracket term for $P=1$, $\alpha=0$, $n=3$, $q=1$, $b=1$, and $k=1$.}
\end{figure}

Applying $\mu=0$, the inversion temperature $T_{i}$ and inversion pressure $P_{i}$ of the BH can be obtained as
\begin{equation}
T_{i} =V\frac{\partial T}{\partial V},
\end{equation}
\begin{equation}
P_{i} =\frac{r^{-2-4\gamma }_{+i}( Aq^{2}r^{4+2n(\gamma -1)}_{+i}(\varepsilon -\omega ) -Br^{6\gamma }_{+i}(2\gamma -1-\varepsilon ))  }
{16\pi (n-\alpha ^{2} )(\alpha ^{2}-1)(\varepsilon -1)}.
\end{equation}
In FIG.3, the effect of the dimensions on the inversion curves is compared while considering $Q$ as a fixed external parameter and disregarding the dilaton field $\alpha$. It is apparent that, similar to most multidimensional BH systems, the inversion temperature increases with the dimension $n$ at low pressures, whereas it decreases with $n$ at high pressures.
\begin{figure}[htbp]
		\centering
		\includegraphics[width=3in,height=2in]{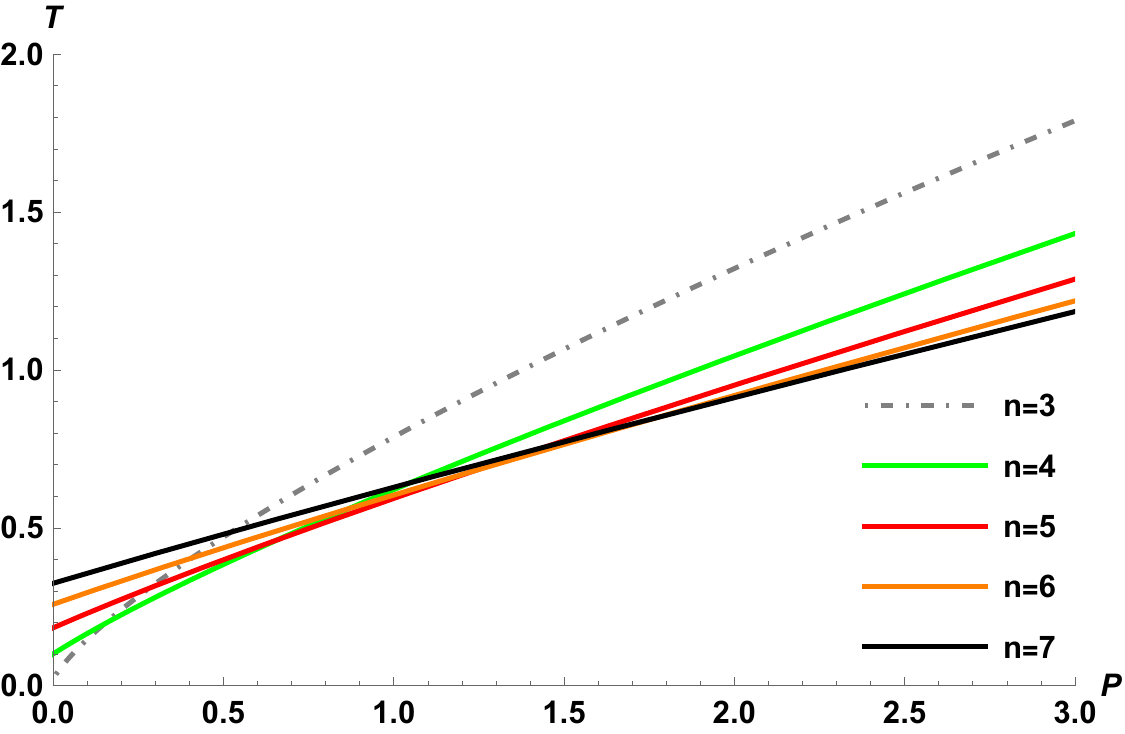}
     \caption{ Effect of the space-time dimensionality on the inversion curve of charged dilatonic black holes for $Q=1$, $\alpha=0$, $b=1$, and $k=1$.}
\end{figure}
\begin{figure}[htbp]
	\centering
	\subfigure[$n=3$,$\alpha=0$]{
    \begin{minipage}[t]{0.32\linewidth}
		\centering
		\includegraphics[width=2in,height=1.2in]{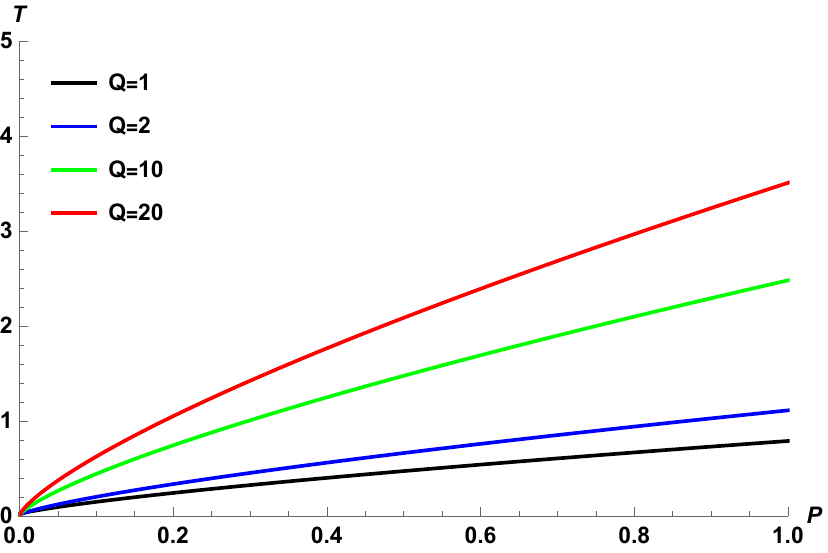}
		\end{minipage}%
            }%
    \subfigure[$n=3$,$\alpha=0.4$]{
    \begin{minipage}[t]{0.32\linewidth}
		\centering
		\includegraphics[width=2in,height=1.2in]{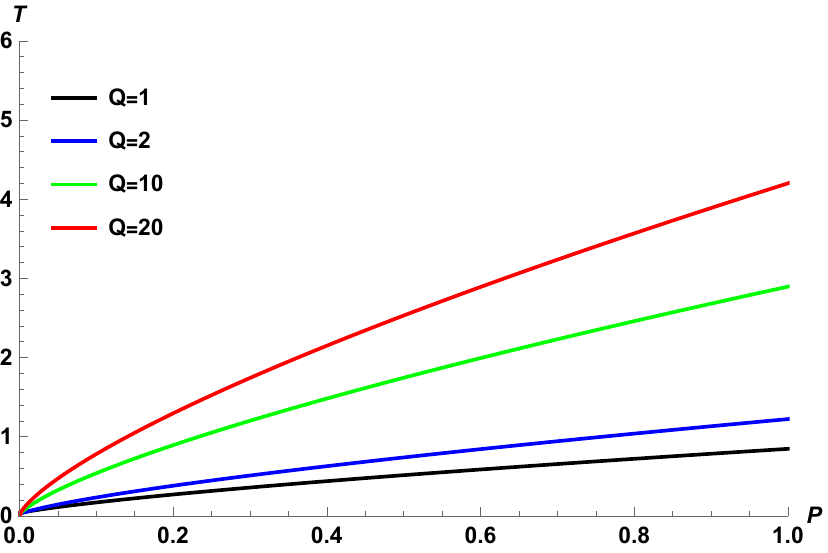}
		\end{minipage}%
            }%
      \subfigure[$n=3$,$\alpha=0.8$]{
    \begin{minipage}[t]{0.32\linewidth}
		\centering
		\includegraphics[width=2in,height=1.2in]{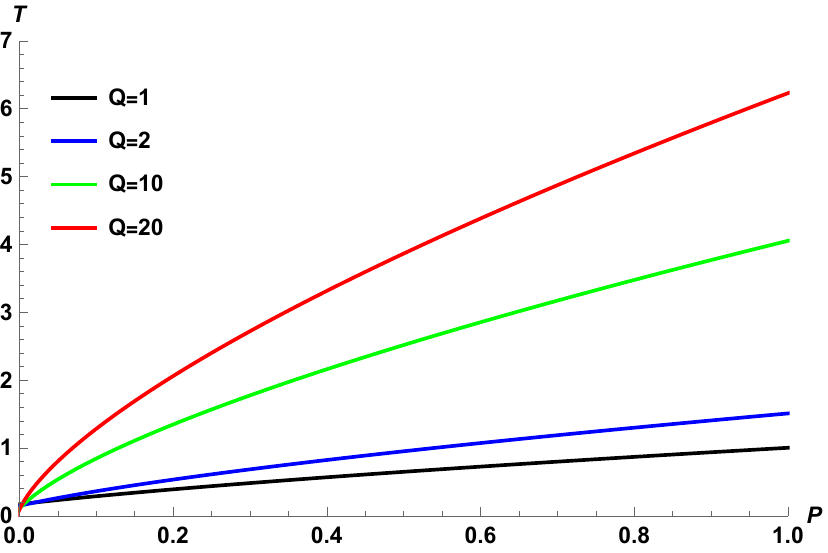}
		\end{minipage}%
            }%

            \subfigure[$n=4$,$\alpha=0$]{
    \begin{minipage}[t]{0.32\linewidth}
		\centering
		\includegraphics[width=2in,height=1.2in]{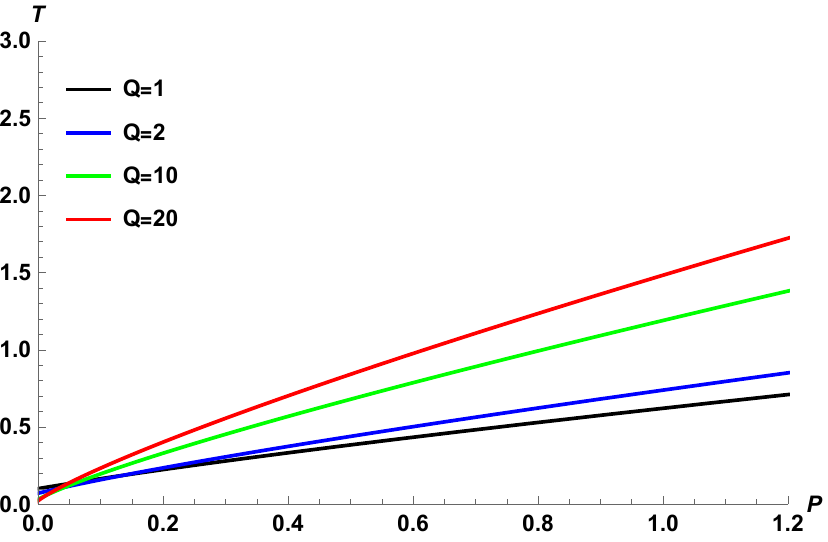}
		\end{minipage}%
            }%
    \subfigure[$n=4$,$\alpha=0.4$]{
    \begin{minipage}[t]{0.32\linewidth}
		\centering
		\includegraphics[width=2in,height=1.2in]{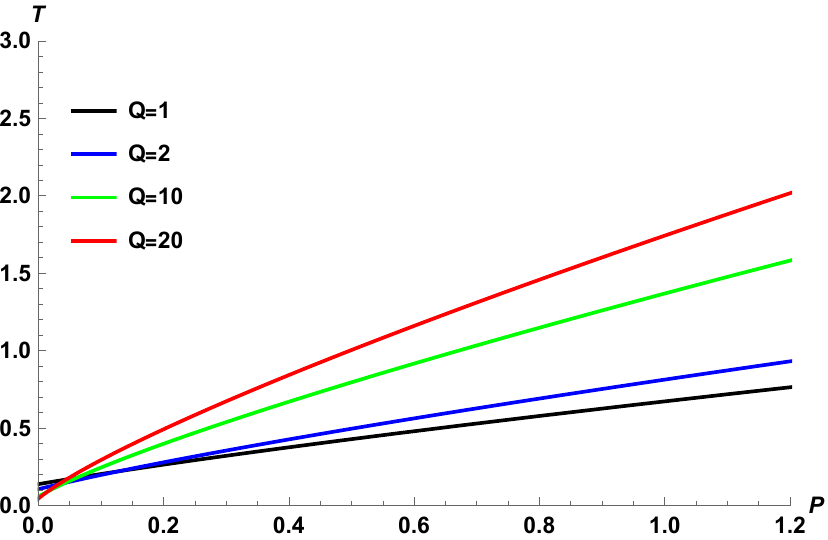}
		\end{minipage}%
            }%
      \subfigure[$n=4$,$\alpha=0.8$]{
    \begin{minipage}[t]{0.32\linewidth}
		\centering
		\includegraphics[width=2in,height=1.2in]{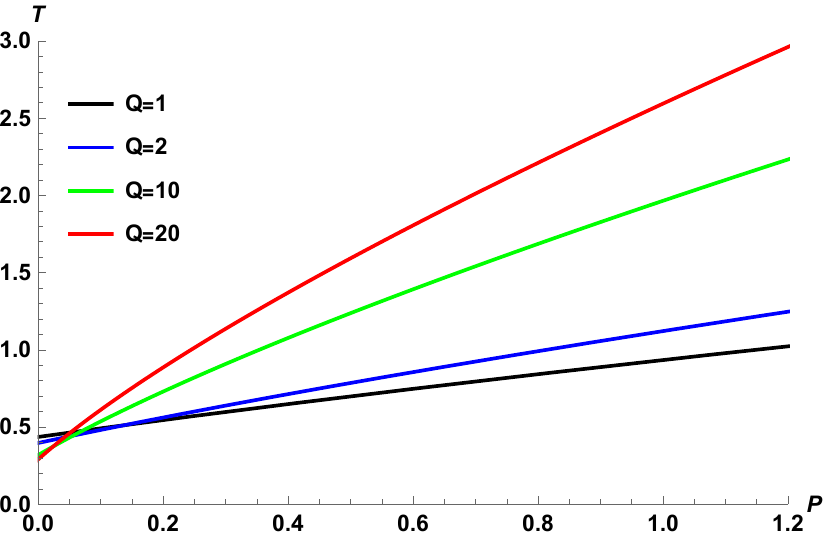}
		\end{minipage}%
            }%

            \subfigure[$n=5$,$\alpha=0$]{
    \begin{minipage}[t]{0.32\linewidth}
		\centering
		\includegraphics[width=2in,height=1.2in]{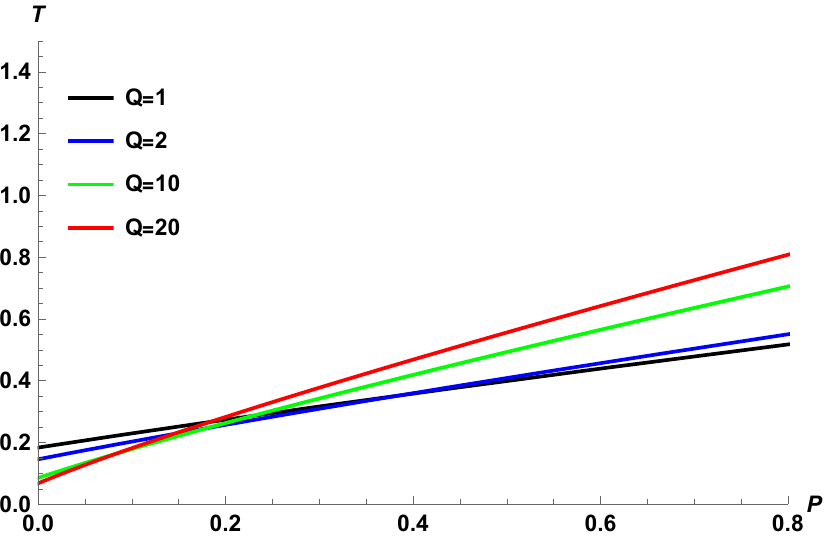}
		\end{minipage}%
            }%
    \subfigure[$n=5$,$\alpha=0.4$]{
    \begin{minipage}[t]{0.32\linewidth}
		\centering
		\includegraphics[width=2in,height=1.2in]{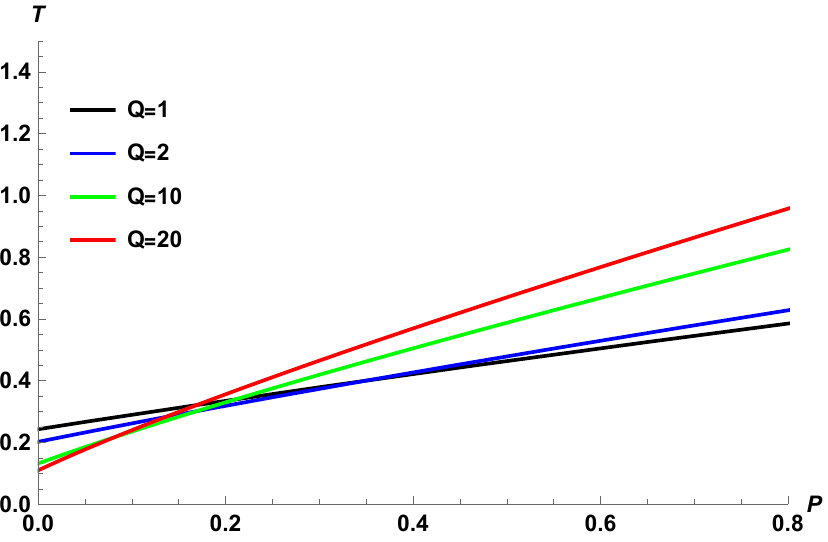}
		\end{minipage}%
            }%
      \subfigure[$n=5$,$\alpha=0.8$]{
    \begin{minipage}[t]{0.32\linewidth}
		\centering
		\includegraphics[width=2in,height=1.2in]{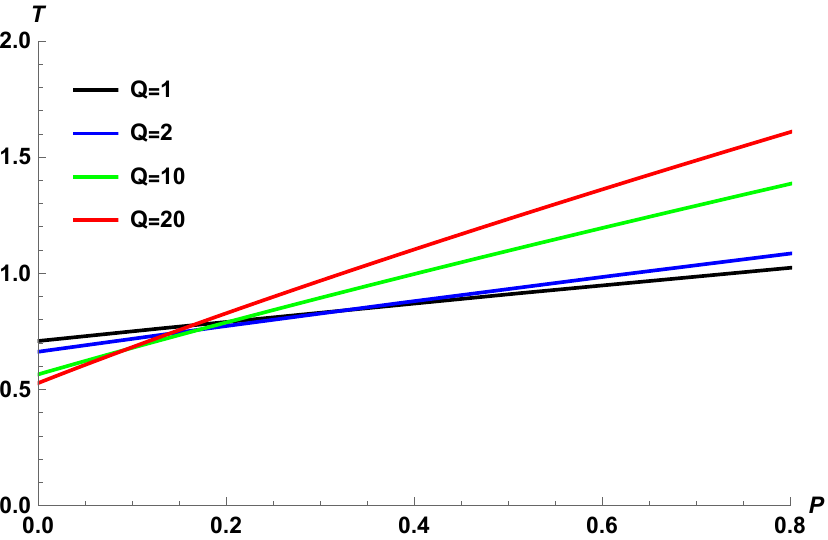}
		\end{minipage}%
            }%

         \subfigure[$n=6$,$\alpha=0$]{
    \begin{minipage}[t]{0.32\linewidth}
		\centering
		\includegraphics[width=2in,height=1.2in]{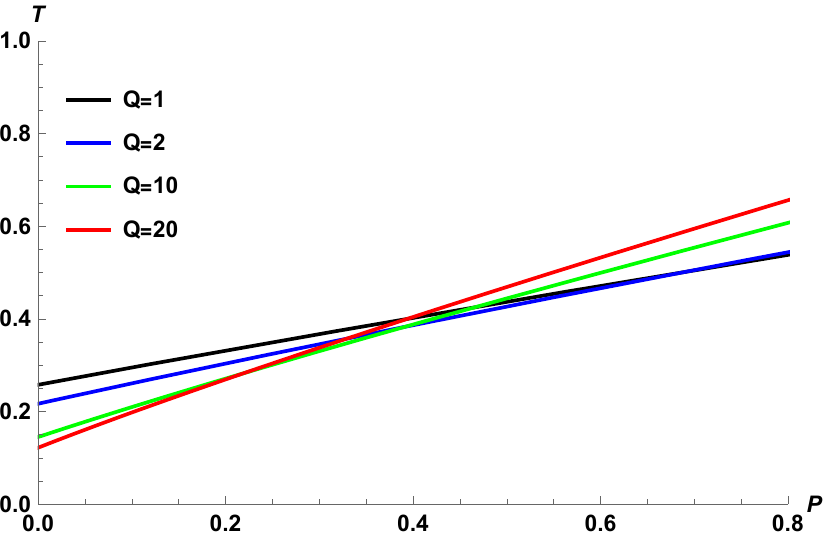}
		\end{minipage}%
            }%
    \subfigure[$n=6$,$\alpha=0.4$]{
    \begin{minipage}[t]{0.32\linewidth}
		\centering
		\includegraphics[width=2in,height=1.2in]{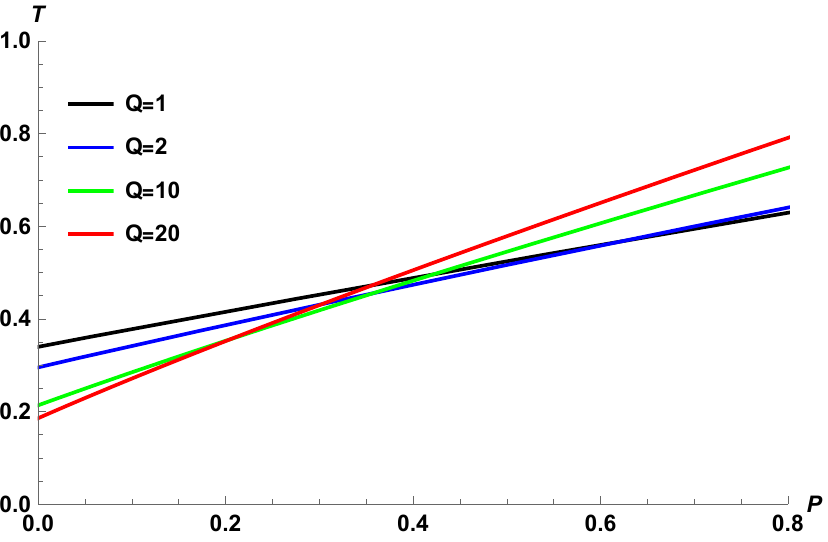}
		\end{minipage}%
            }%
      \subfigure[$n=6$,$\alpha=0.8$]{
    \begin{minipage}[t]{0.32\linewidth}
		\centering
		\includegraphics[width=2in,height=1.2in]{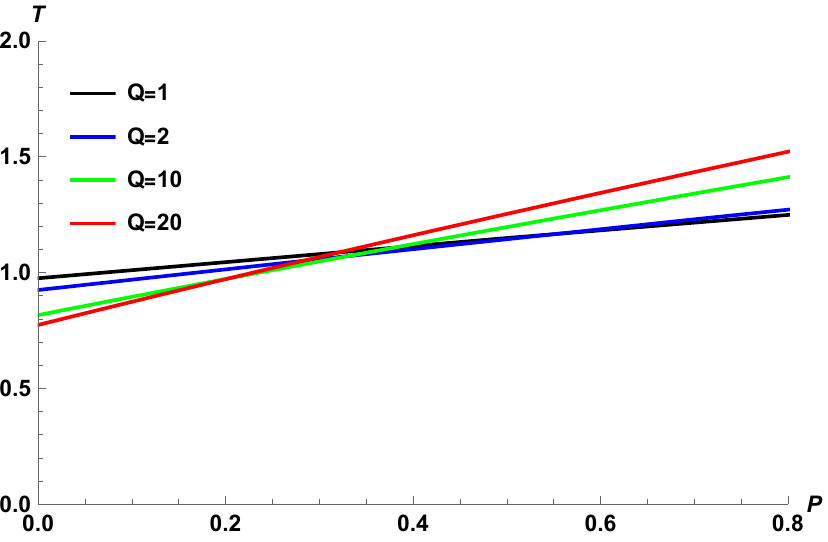}
		\end{minipage}%
            }%
     \centering
     \caption{Inversion curves of charged dilatonic black holes for $b=1$ and $k=1$.}
\end{figure}

From FIG. 4, we can observe the influence of the charge $Q$ on the inversion curve in the T-P plane. In contrast to the effect of dimensions (see FIG. 3), at low pressures, the inverse temperature decreases as the charge $Q$ increases, whereas the opposite result is observed at high pressures. Next, considering the subfigures in the first column of FIG. 4 (see FIG. 4(a), FIG. 4(d), FIG. 4(g), and FIG. 4(j)),  we can observe that as $n$ increases, the high and low pressure cut-off points move to the upper right. This result also holds for other values of the parameter $\alpha$. Based on the fourth row of the figure, we analyzed the effect of the parameter $\alpha$ on the inversion curve (see FIG. 4(j), FIG. 4(k) and FIG. 4(l)) and discovered that as $\alpha$ increases, the high and low pressure cut-off points move to the upper left; this phenomenon is not evident at lower dimensions; however, the high and low pressure cut-off points still move upward. This indicates that the influence of $\alpha$ on the BH may be more apparent in higher dimensional space-time. In addition, an increase in the dimension $n$ or parameter $\alpha$ leads to an increase in $T_{min}$, corresponding to the zero inversion pressure $P_{i}=0$.

Furthermore, it is of great significance to study the ratio between the minimum inversion temperature $T_{min}$  and the critical temperature $T_{c}$.  Note that the equations for $r_{min}$ and $T_{min}$ can be obtained by ensuring $P_{i}=0$.
\begin{equation}
r^{-2-4\gamma }_{min}( 4A \pi^{2-n}Q^{2}\Gamma (\frac{n}{2} )^{2}r^{4+2n(\gamma -1)}_{min}(\varepsilon -\omega ) -Br^{6\gamma }_{min}(2\gamma -1-\varepsilon ))=0,
\end{equation}
\begin{equation}
T_{min} =(1+\alpha ^{2} )r_{min} [\frac{-4A \omega \pi ^{2-n}Q^{2} r_{min}^{2+2n(\gamma -1)-4\gamma }\Gamma (\frac{n}{2} )^{2}   -Br_{min}^{2\gamma -2}(2\gamma -1) }{4\pi (n-1)(\alpha ^{2}-1 )(n+\alpha ^{2} )\varepsilon} ].
\end{equation}
As the minimum radius in Eq.(25) contains a complex index, it is difficult to obtain explicit expressions for the ratio $T_{min}/T_{c}$; however, we can still perform some numerical analysis. By combining Eqs.(14), (25) and (26), the numerical results of the ratio $T_{min}/T_{c}$ are shown in FIG. 5. It is evident that the ratio $T_{min}/T_{c}$ is independent of the charge; however, an increase in the dilaton parameter $\alpha$ causes the ratio to increase. This phenomenon can be observed at any dimension. For small values of $\alpha$ (including $\alpha=0$),  the ratio decreases with an increase in the dimension $n$, and under certain conditions, the ratio is less than 1/2. A case corresponding to $\alpha=0$ is shown in the data in Table I. As the parameter $\alpha$ approaches 1, when $n$ increases, the ratio $T_{min}/T_{c}$ increases, and it is always greater than 1/2. In fact, we can also perform an analytical investigation on the ratio $T_{min}/T_{c}$ under some special conditions; for instance, for $n=3$ and $\alpha=0$, $r_{min}$  has the following root
\begin{equation}
r_{min}=-\frac{\sqrt{\frac{3}{2} } Q}{\sqrt{k} },r_{min}=\frac{\sqrt{\frac{3}{2} } Q}{\sqrt{k} }.
\end{equation}
To make the ratio $T_{min}/T_{c}$ positive, we consider $r_{min}$ to be a physically positive root and substitute the root into Eq. (26) and obtain
\begin{equation}
T_{min}=\frac{k^{\frac{3}{2} } } {6\sqrt{6} \pi Q },      T_{c}=\frac{\sqrt{\frac{1}{Q^{2} } } }{3\sqrt{6}\pi(\frac{1}{k} )^{\frac{3}{2} }  } .
\end{equation}
We obtain $T_{min}/T_{c}=1/2$, in agreement with the result reported in Ref \cite{J1}. The other physical quantities, except for the dilaton field $\alpha$, at various dimensions are summarized in Table II.
\begin{figure}[htbp]
	\centering
	\subfigure[$n=3$]{
    \begin{minipage}[t]{0.32\linewidth}
		\centering
		\includegraphics[width=2in,height=1.2in]{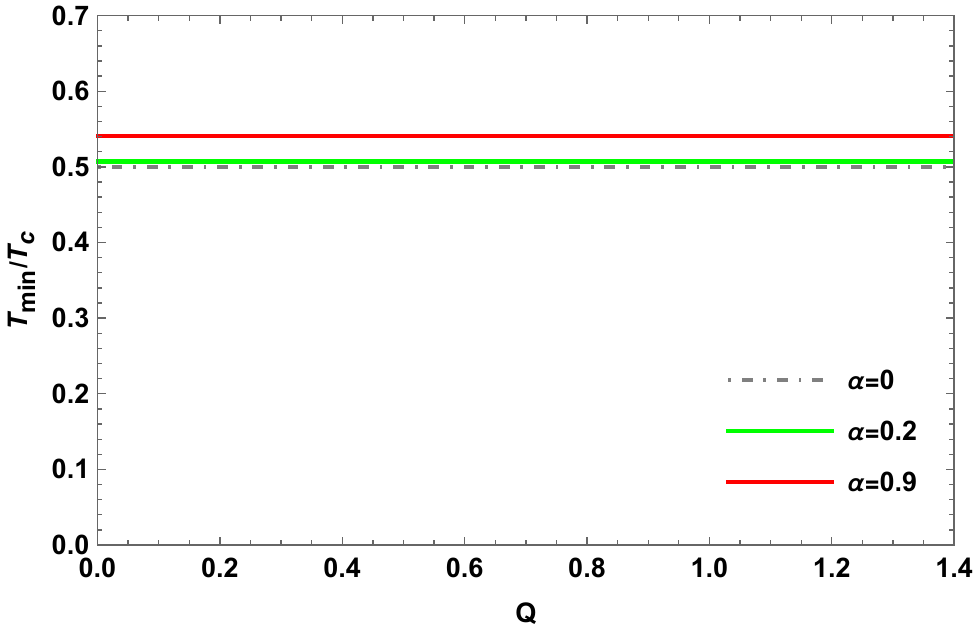}
		\end{minipage}%
            }%
    \subfigure[$n=4$]{
    \begin{minipage}[t]{0.32\linewidth}
		\centering
		\includegraphics[width=2in,height=1.2in]{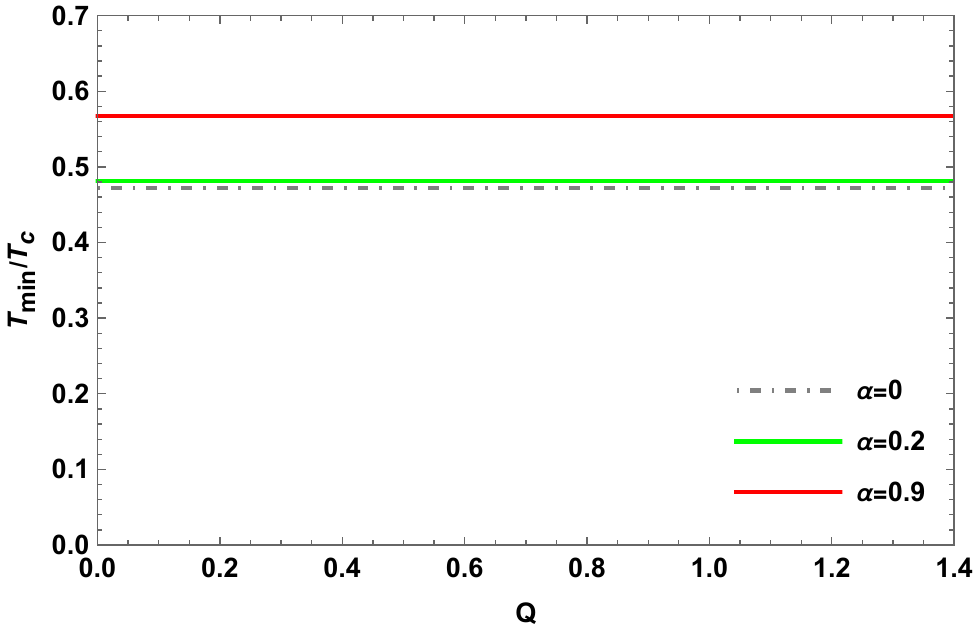}
		\end{minipage}%
            }%
      \subfigure[$n=5$]{
    \begin{minipage}[t]{0.32\linewidth}
		\centering
		\includegraphics[width=2in,height=1.2in]{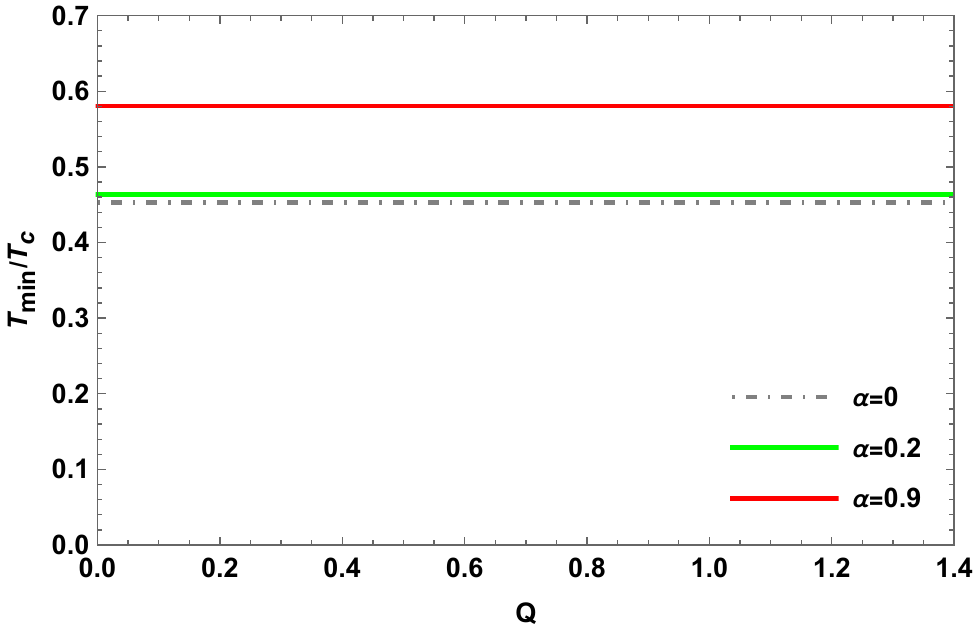}
		\end{minipage}%
            }%

            \subfigure[$n=6$]{
    \begin{minipage}[t]{0.32\linewidth}
		\centering
		\includegraphics[width=2in,height=1.2in]{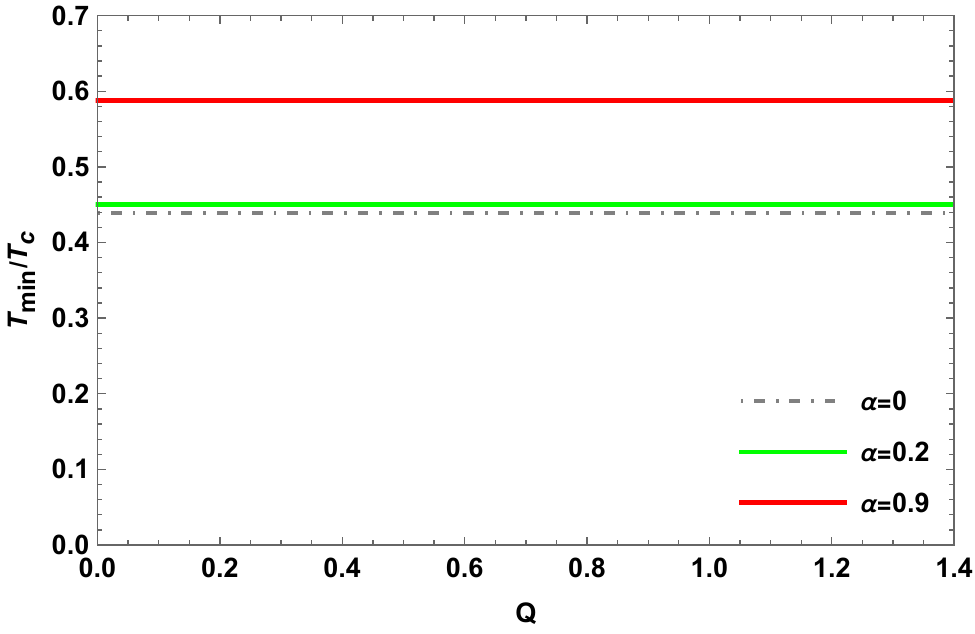}
		\end{minipage}%
            }%
    \subfigure[$n=7$]{
    \begin{minipage}[t]{0.32\linewidth}
		\centering
		\includegraphics[width=2in,height=1.2in]{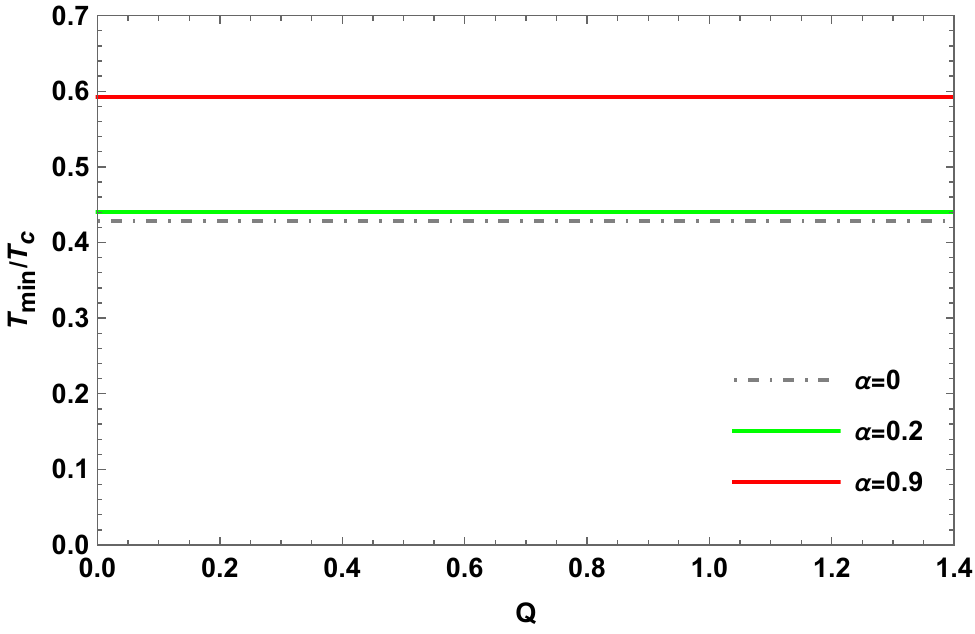}
		\end{minipage}%
            }%
     \centering
     \caption{Ratio $T_{min}/T_{c}$ at different dimensions with $b=1$ and $k=1$.}
\end{figure}
\begin{equation}
\begin{aligned}
&\text { Table I: Ratio $\frac{T_{min}}{T_{c}}$ at various dimensions.}\\
& \begin{array}{c|l|l|l|l|l}\setlength
\hline \hline d  (\alpha=0) & 3 & 4 & 5 & 6 & 7 \\
\hline \frac{T_{min}}{T_{c}} & 0.5 & 0.471957 & 0.452802& 0.438933 & 0.428377 \\
\hline \hline
\end{array}
\end{aligned}
\end{equation}
\begin{equation}
\begin{aligned}
&\text { Table II: Other physical quantities }\left(r_{min}, T_{min} \text { and } T_{c}\right) \text { for various dimensions. }\\
&\begin{array}{c|l|l|l}
\hline \hline d (\alpha=0)& r_{min} & T_{min} & T_{c} \\
\hline 4 & \frac{12Q^{2}-5\pi r_{min}^{4}k  }{8\pi ^{3}r_{min}^{6} } =0 & \frac{1}{4\pi } (\frac{2k}{r_{min} } -\frac{8Q^{2} }{3\pi^{2}r_{min}^{5} } ) & \frac{2\sqrt{2} (\frac{1}{Q^{2} } )^{\frac{1}{4} }}{5\times 5^{1/4} \sqrt{\pi}(\frac{1}{k} )^{\frac{5}{4} }  } \\
\hline 5 & \frac{9(3Q^{2}-4\pi^{2}  r_{min}^{6}k  )}{32\pi ^{3}r_{min}^{8} } =0 & \frac{1}{4\pi } (\frac{3k}{r_{min} } -\frac{9Q^{2} }{8\pi^{2}r_{min}^{7} } )  & \frac{3(\frac{2}{7} )^{1/6}3^{5/6} (\frac{1}{Q^{2} } )^{1/6 }}{7\pi^{2/3}(\frac{1}{k} )^{7/6 }  } \\
\hline 6 & \frac{6Q^{2}}{\pi ^{5}r_{min}^{10} } -\frac{7k}{4\pi r_{min}^{2} } =0 & \frac{1}{4\pi } (\frac{4k}{r_{min} } -\frac{32Q^{2} }{5\pi^{4}r_{min}^{9} } )  & \frac{8\times 2^{5/8} (\frac{1}{Q^{2} } )^{1/8 }}{9\times 3^{1/4}\sqrt{\pi} (\frac{1}{k} )^{9/8 }  }  \\
\hline 7 & \frac{675Q^{2}}{128\pi ^{5}r_{min}^{12} } -\frac{5k}{2\pi r_{min}^{2} } =0 & \frac{1}{4\pi } (\frac{5k}{r_{min} } -\frac{75Q^{2} }{16\pi^{4}r_{min}^{11} } )  & \frac{5\times 2^{3/10}\times5^{9/10} (\frac{1}{Q^{2} } )^{1/10 }}{11\times 3^{1/5}\times11^{1/10}\pi^{3/5} (\frac{1}{k} )^{11/10 }  }  \\
\hline \hline
\end{array}
\end{aligned}
\end{equation}
\begin{figure}[htbp]
	\centering
	\subfigure[$n=3$,$\alpha=0$]{
    \begin{minipage}[t]{0.32\linewidth}
		\centering
		\includegraphics[width=2in,height=1.2in]{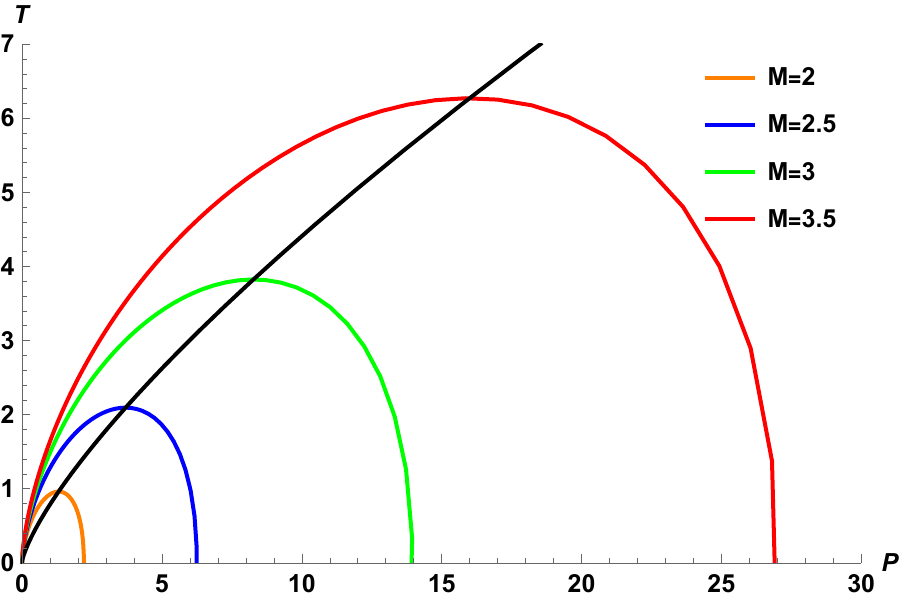}
		\end{minipage}%
            }%
    \subfigure[$n=3$,$\alpha=0.4$]{
    \begin{minipage}[t]{0.32\linewidth}
		\centering
		\includegraphics[width=2in,height=1.2in]{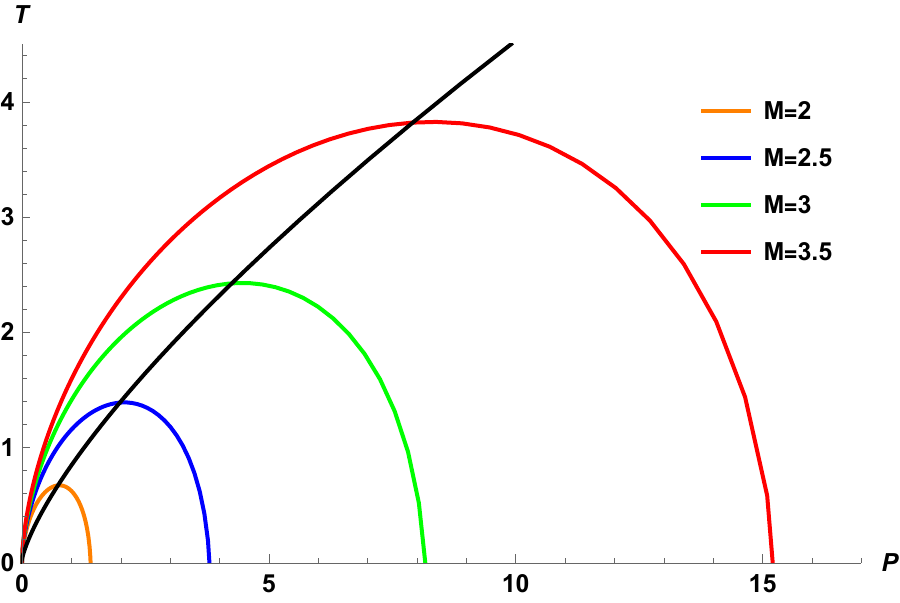}
		\end{minipage}%
            }%
      \subfigure[$n=3$,$\alpha=0.8$]{
    \begin{minipage}[t]{0.32\linewidth}
		\centering
		\includegraphics[width=2in,height=1.2in]{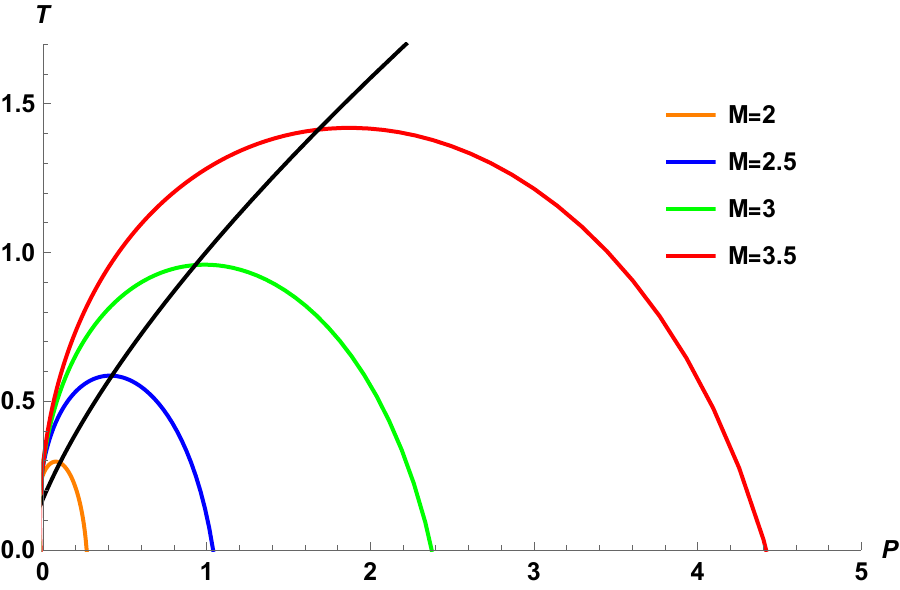}
		\end{minipage}%
            }%

            \subfigure[$n=4$,$\alpha=0$]{
    \begin{minipage}[t]{0.32\linewidth}
		\centering
		\includegraphics[width=2in,height=1.2in]{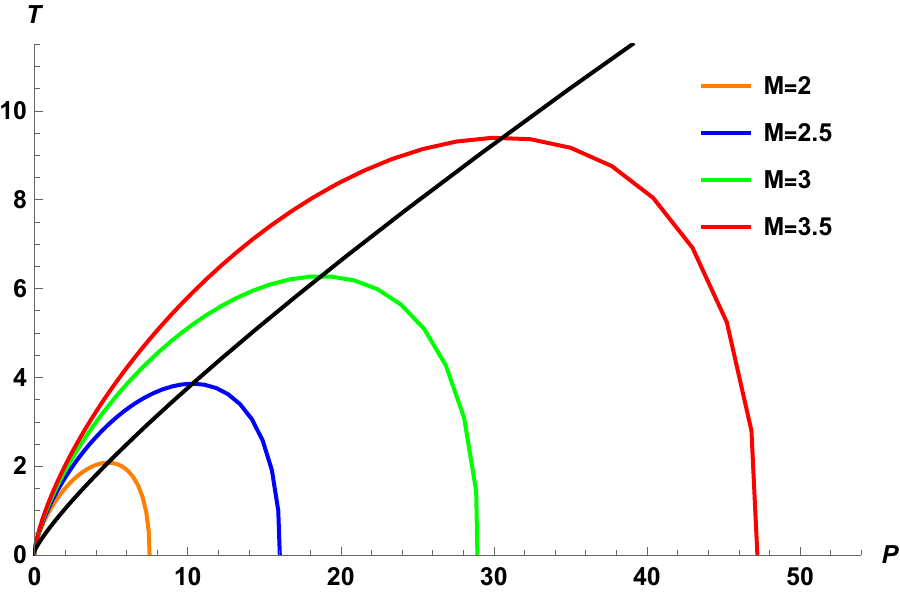}
		\end{minipage}%
            }%
    \subfigure[$n=4$,$\alpha=0.4$]{
    \begin{minipage}[t]{0.32\linewidth}
		\centering
		\includegraphics[width=2in,height=1.2in]{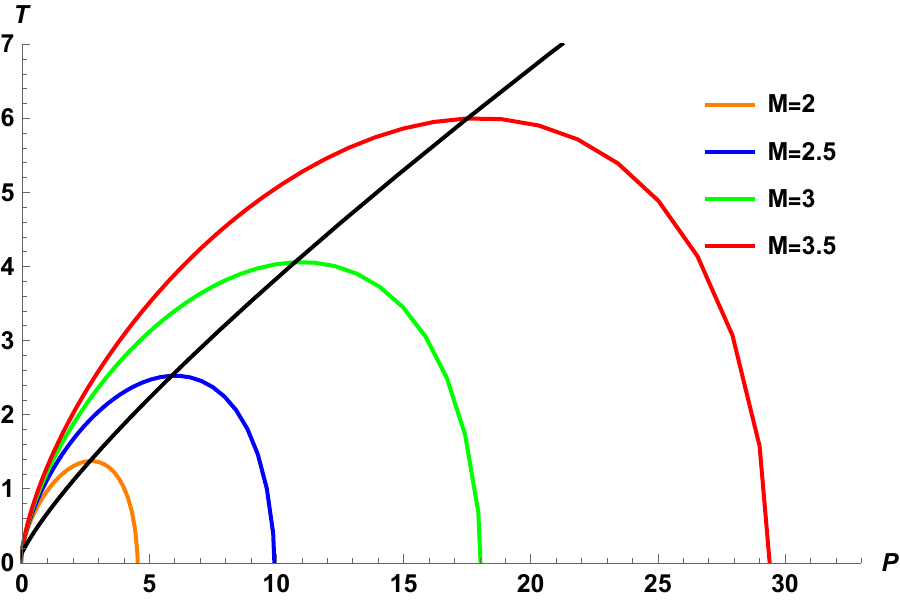}
		\end{minipage}%
            }%
      \subfigure[$n=4$,$\alpha=0.8$]{
    \begin{minipage}[t]{0.32\linewidth}
		\centering
		\includegraphics[width=2in,height=1.2in]{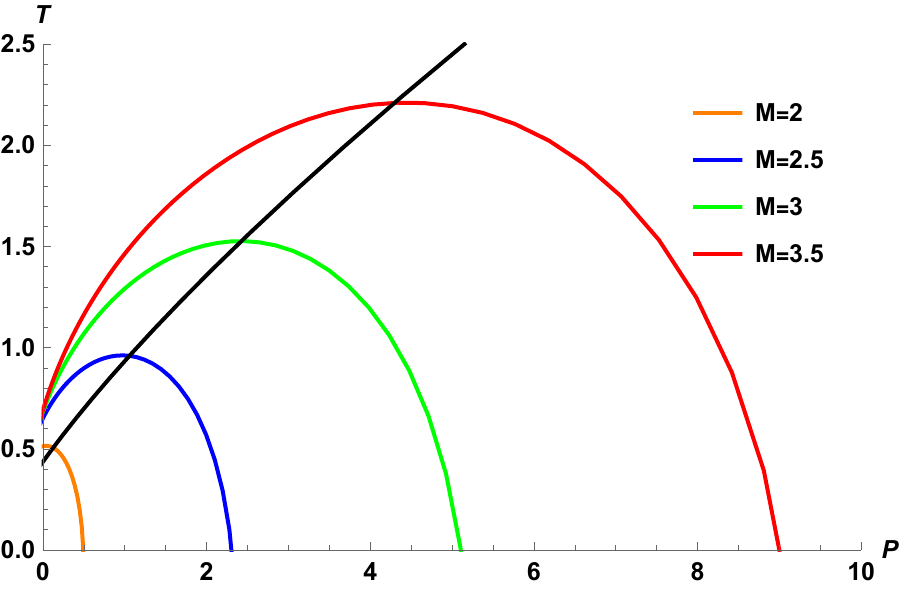}
		\end{minipage}%
            }%

            \subfigure[$n=5$,$\alpha=0$]{
    \begin{minipage}[t]{0.32\linewidth}
		\centering
		\includegraphics[width=2in,height=1.2in]{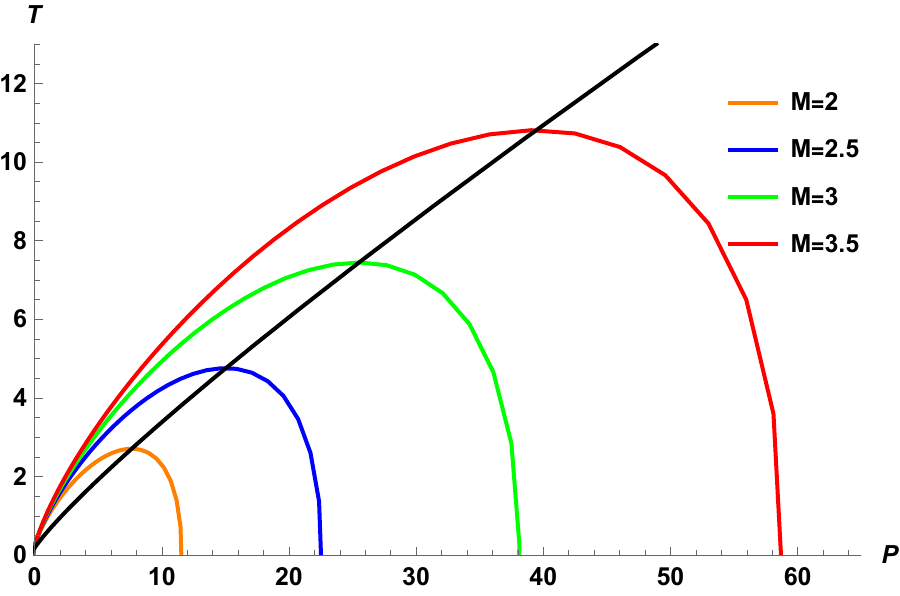}
		\end{minipage}%
            }%
    \subfigure[$n=5$,$\alpha=0.4$]{
    \begin{minipage}[t]{0.32\linewidth}
		\centering
		\includegraphics[width=2in,height=1.2in]{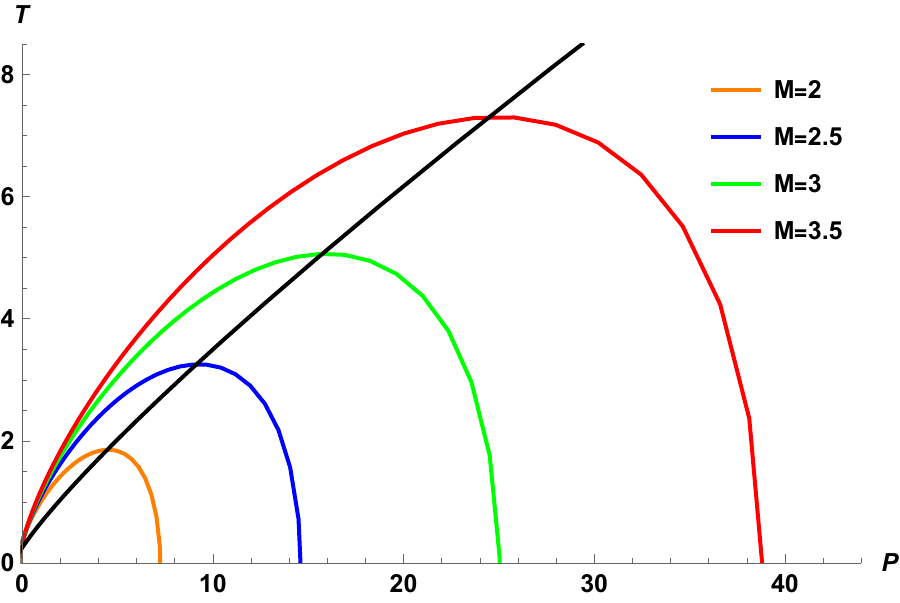}
		\end{minipage}%
            }%
      \subfigure[$n=5$,$\alpha=0.8$]{
    \begin{minipage}[t]{0.32\linewidth}
		\centering
		\includegraphics[width=2in,height=1.2in]{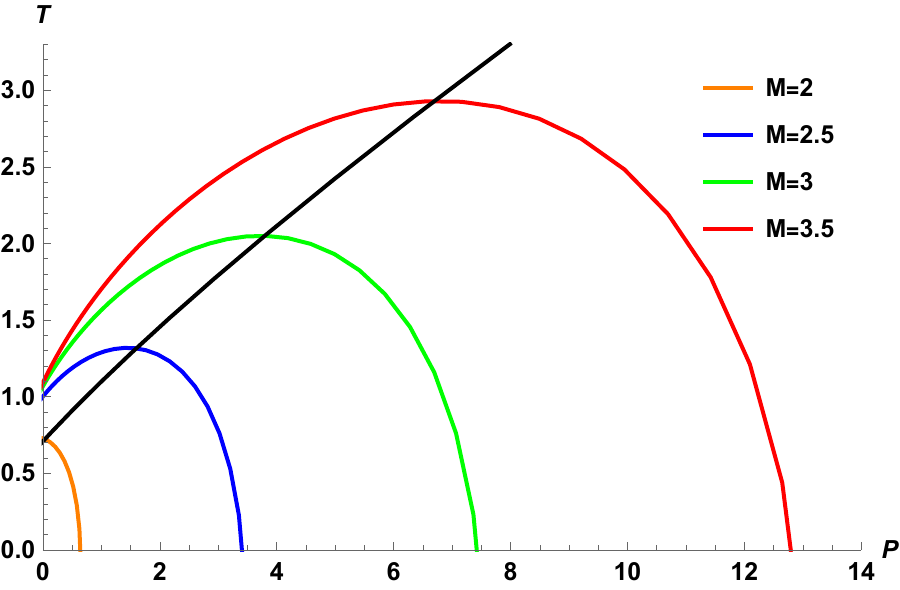}
		\end{minipage}%
            }%

         \subfigure[$n=6$,$\alpha=0$]{
    \begin{minipage}[t]{0.32\linewidth}
		\centering
		\includegraphics[width=2in,height=1.2in]{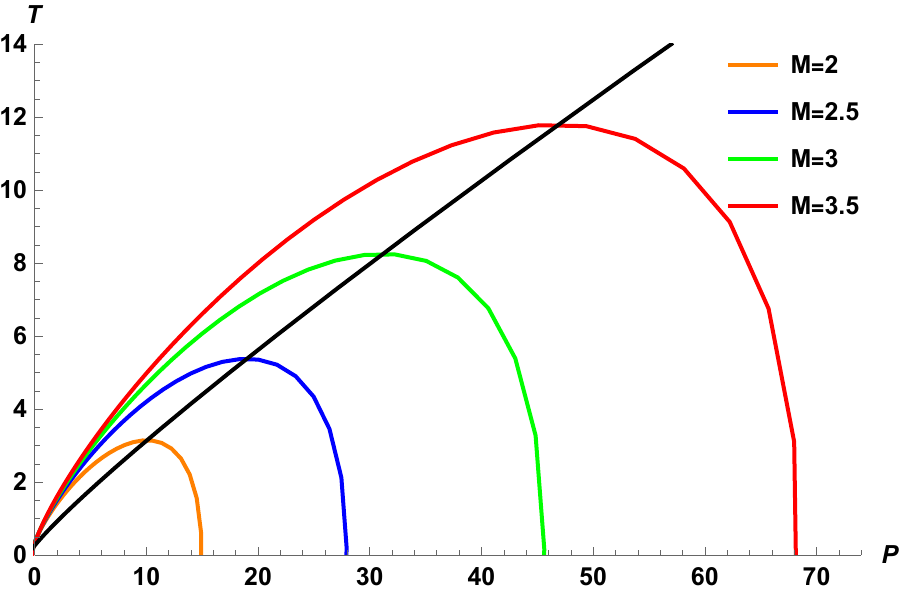}
		\end{minipage}%
            }%
    \subfigure[$n=6$,$\alpha=0.4$]{
    \begin{minipage}[t]{0.32\linewidth}
		\centering
		\includegraphics[width=2in,height=1.2in]{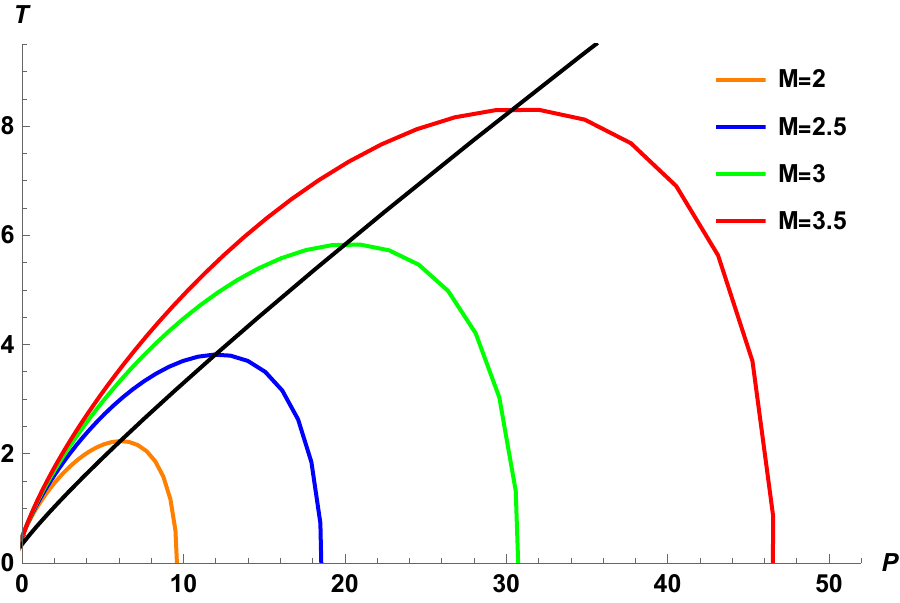}
		\end{minipage}%
            }%
      \subfigure[$n=6$,$\alpha=0.8$]{
    \begin{minipage}[t]{0.32\linewidth}
		\centering
		\includegraphics[width=2in,height=1.2in]{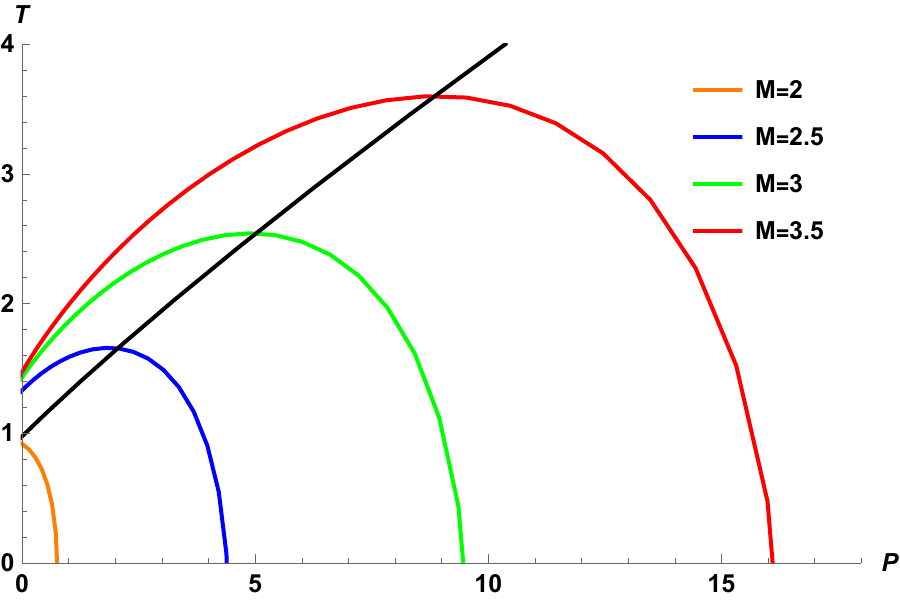}
		\end{minipage}%
            }%
     \centering
     \caption{Isenthalpic curves and inversion curves for $Q=1$, $b=1$, and $k=1$.}
\end{figure}
\begin{figure}[htbp]
	\centering
	\subfigure[$n=3$,$\alpha=0$]{
    \begin{minipage}[t]{0.32\linewidth}
		\centering
		\includegraphics[width=2in,height=1.2in]{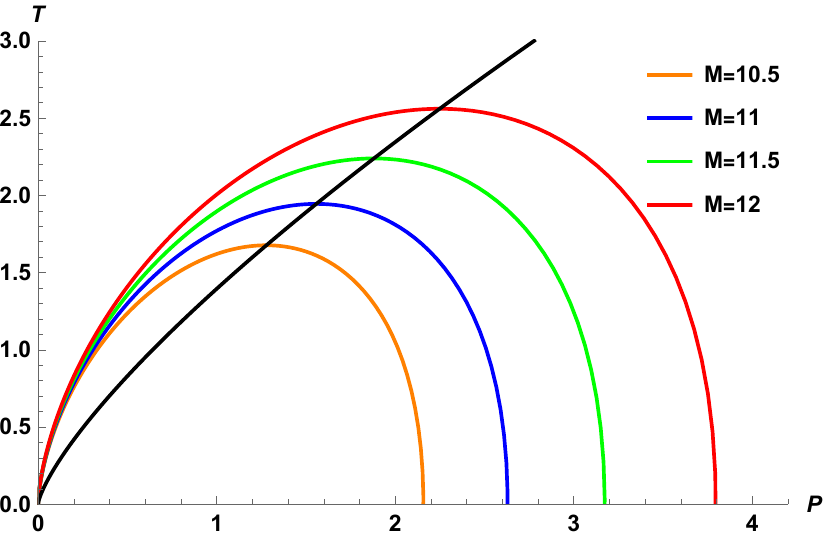}
		\end{minipage}%
            }%
    \subfigure[$n=3$,$\alpha=0.4$]{
    \begin{minipage}[t]{0.32\linewidth}
		\centering
		\includegraphics[width=2in,height=1.2in]{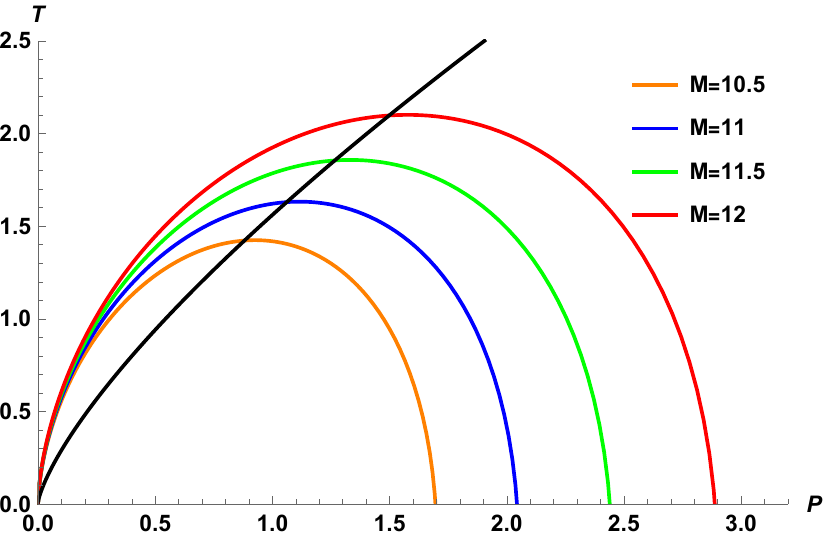}
		\end{minipage}%
            }%
      \subfigure[$n=3$,$\alpha=0.8$]{
    \begin{minipage}[t]{0.32\linewidth}
		\centering
		\includegraphics[width=2in,height=1.2in]{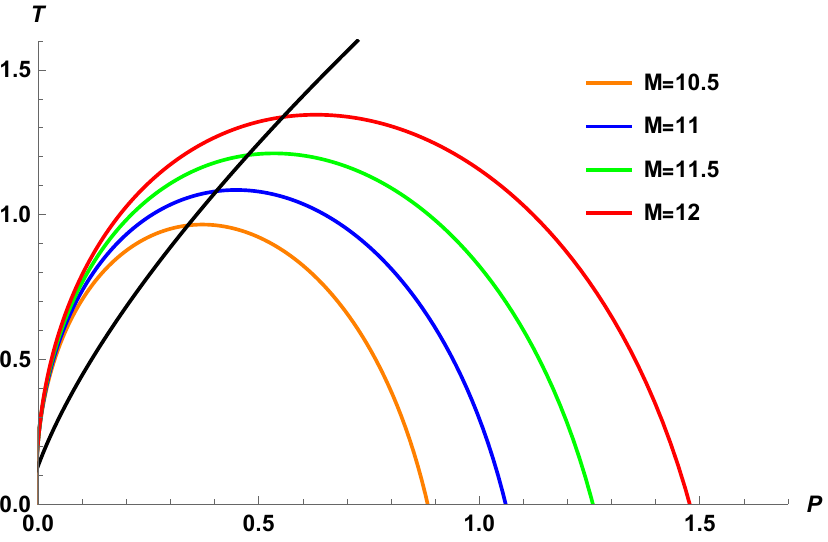}
		\end{minipage}%
            }%

            \subfigure[$n=4$,$\alpha=0$]{
    \begin{minipage}[t]{0.32\linewidth}
		\centering
		\includegraphics[width=2in,height=1.2in]{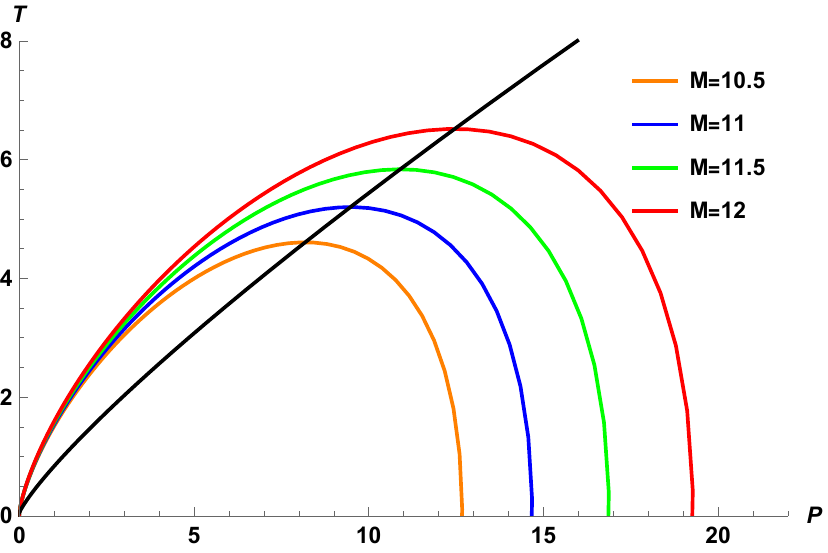}
		\end{minipage}%
            }%
    \subfigure[$n=4$,$\alpha=0.4$]{
    \begin{minipage}[t]{0.32\linewidth}
		\centering
		\includegraphics[width=2in,height=1.2in]{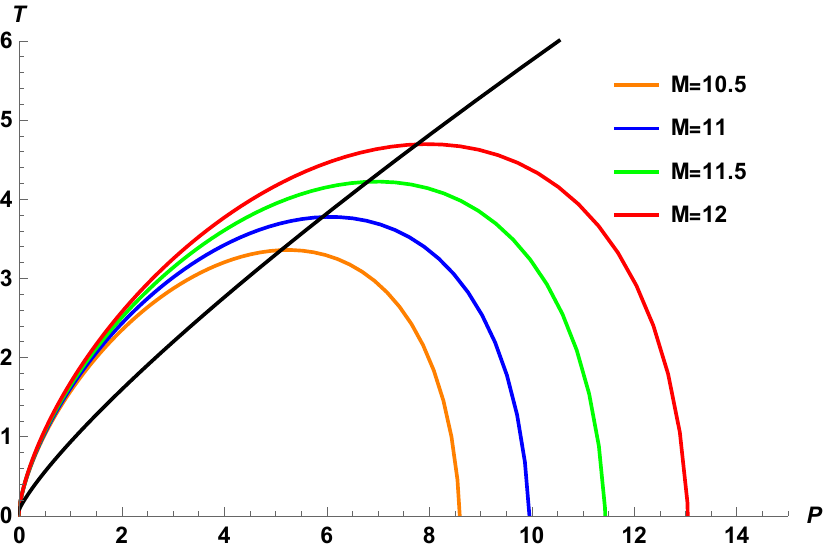}
		\end{minipage}%
            }%
      \subfigure[$n=4$,$\alpha=0.8$]{
    \begin{minipage}[t]{0.32\linewidth}
		\centering
		\includegraphics[width=2in,height=1.2in]{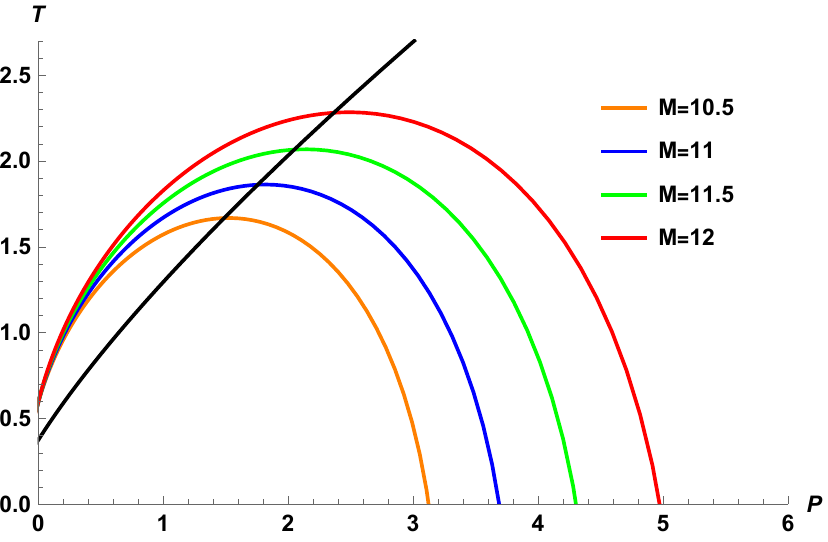}
		\end{minipage}%
            }%

            \subfigure[$n=5$,$\alpha=0$]{
    \begin{minipage}[t]{0.32\linewidth}
		\centering
		\includegraphics[width=2in,height=1.2in]{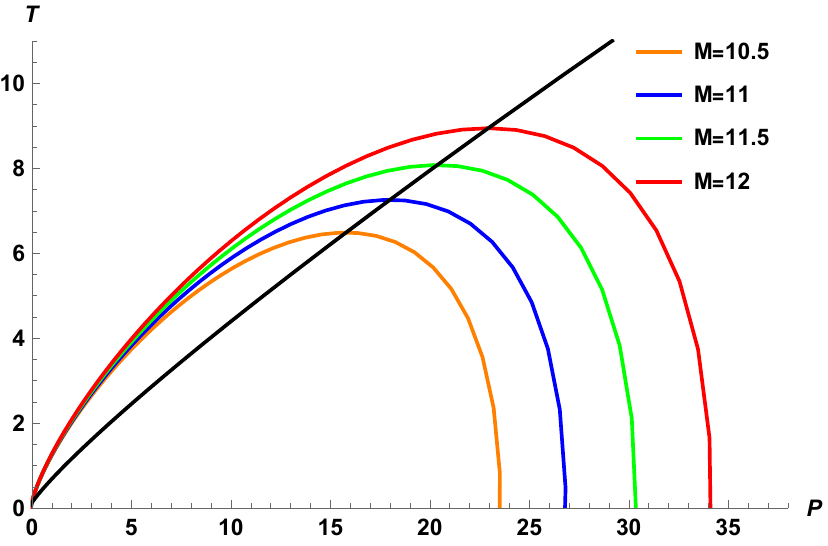}
		\end{minipage}%
            }%
    \subfigure[$n=5$,$\alpha=0.4$]{
    \begin{minipage}[t]{0.32\linewidth}
		\centering
		\includegraphics[width=2in,height=1.2in]{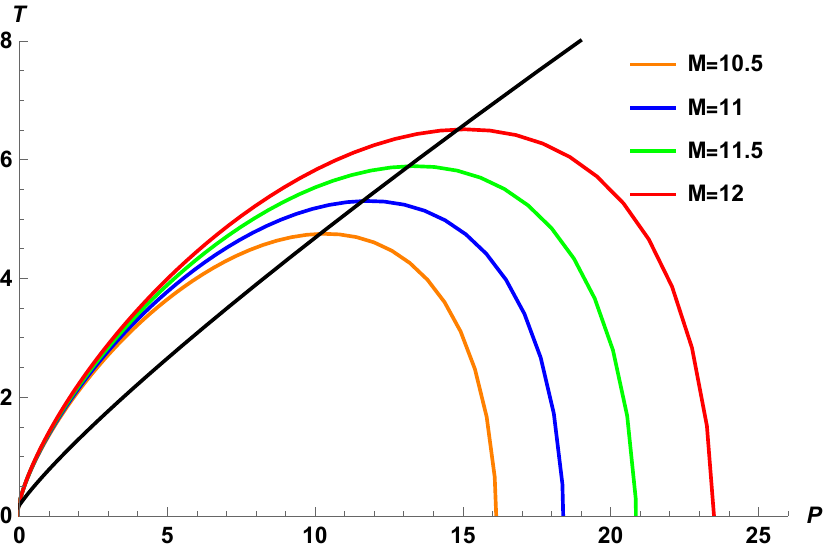}
		\end{minipage}%
            }%
      \subfigure[$n=5$,$\alpha=0.8$]{
    \begin{minipage}[t]{0.32\linewidth}
		\centering
		\includegraphics[width=2in,height=1.2in]{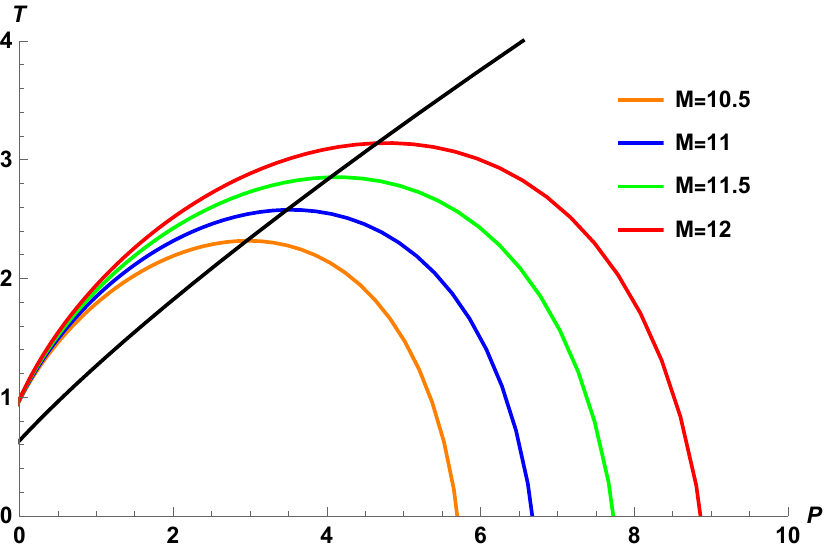}
		\end{minipage}%
            }%

         \subfigure[$n=6$,$\alpha=0$]{
    \begin{minipage}[t]{0.32\linewidth}
		\centering
		\includegraphics[width=2in,height=1.2in]{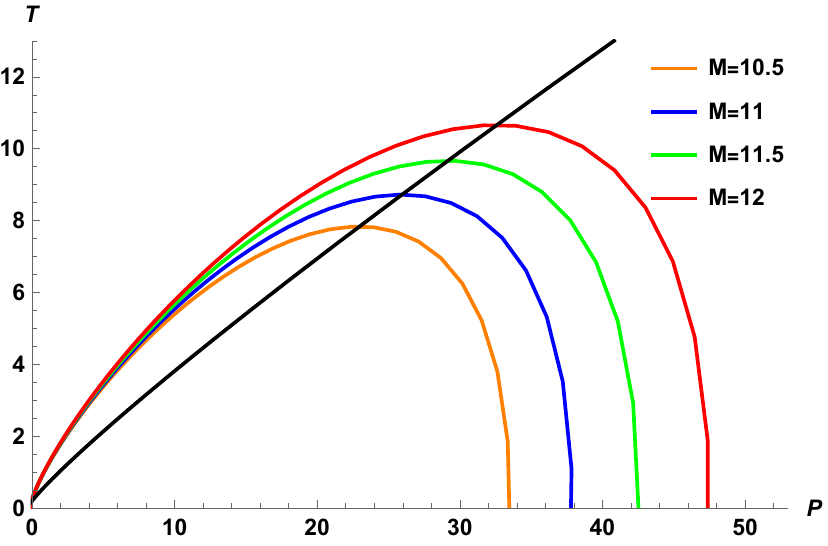}
		\end{minipage}%
            }%
    \subfigure[$n=6$,$\alpha=0.4$]{
    \begin{minipage}[t]{0.32\linewidth}
		\centering
		\includegraphics[width=2in,height=1.2in]{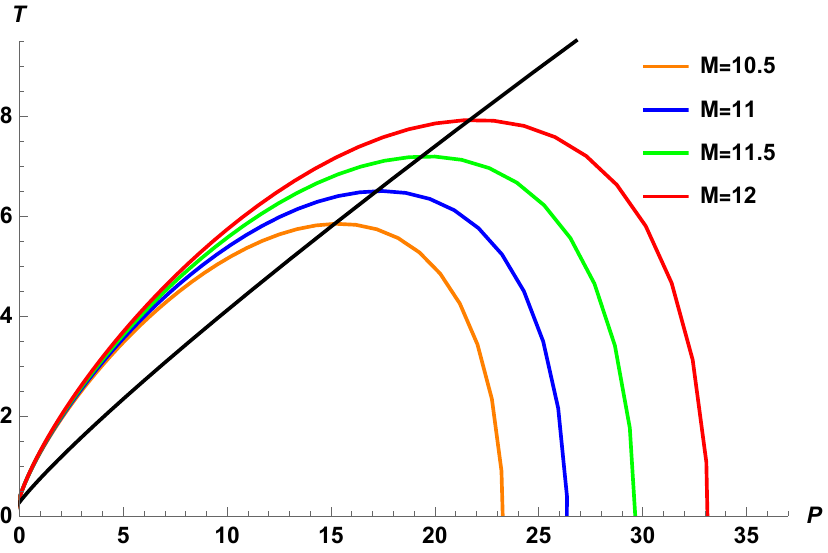}
		\end{minipage}%
            }%
      \subfigure[$n=6$,$\alpha=0.8$]{
    \begin{minipage}[t]{0.32\linewidth}
		\centering
		\includegraphics[width=2in,height=1.2in]{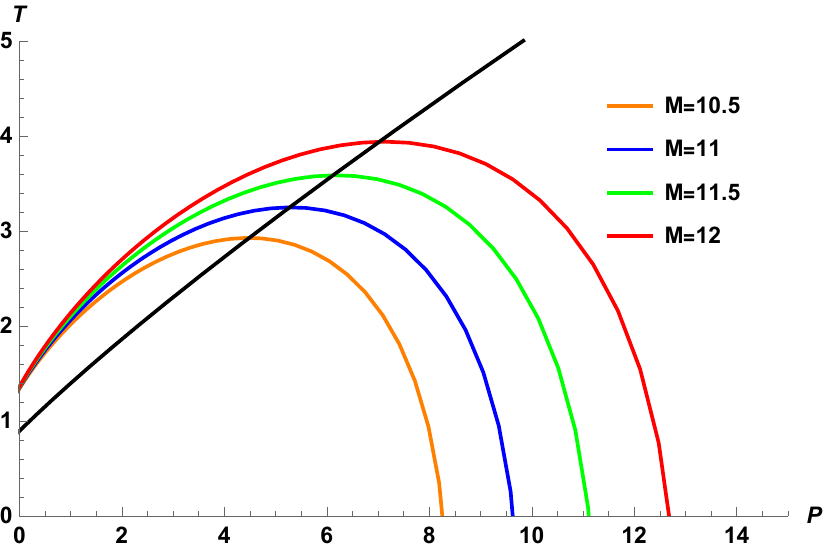}
		\end{minipage}%
            }%
     \centering
     \caption{Isenthalpic curves and inversion curves for $Q=10$, $b=1$, and $k=1$.}
\end{figure}

\begin{figure}[htbp]
		\centering
		\includegraphics[width=3in,height=2in]{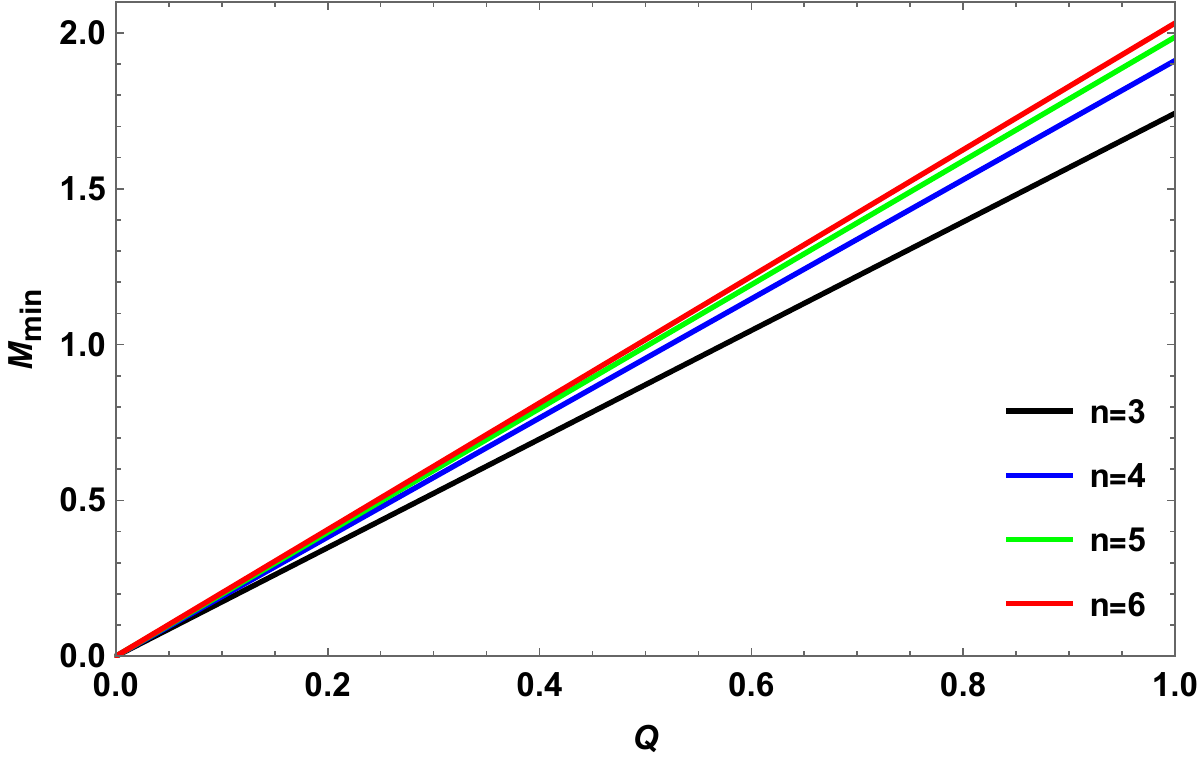}
     \caption{ Minimum inversion mass $M_{min}$ at different dimensions with $\alpha=0.8$, $b=1$, and $k=1$.}
\end{figure}

As J-T expansion is an isenthalpy process, we further analyzed the isenthalpy curves. By using Eqs(4), (5), (7), (23), and (24), we plot the isenthalpy and inversion curves in plane $T-P$ by fixing the charge Q. To avoid the appearance of a bare singularity, we set $M>Q$.  Figures 6 and 7 indicate that the inversion curves (black curves) coincide with the maximum point of the isenthalpic curves (colored curves). We can also identify a region where the constant enthalpy curve has a higher slope than the inverse curve, that is, the region where cooling occurs.  The sign of the slope of the isenthalpic curves changes under the inversion curves, implying that this region presents signs of warming. Therefore, the boundary between the heating and cooling regions of the BH depends on the inversion curve. Upon close examination, we discover that the isenthalpic curves expand to the right as $M$ increases, and the cooling-heating critical points shift upward accordingly.  As $n$ increases, the curve tends to expand at higher pressures. By comparing FIG. 6 and  FIG. 7, it is evident that as Q increases, the curves tend to move toward the left, and the cooling-heating critical point changes. Next, we substitute Eq. (25) into Eq. (5) and obtain the minimum inversion mass $M_{min}$. We present the behavior of the mass $M_{min}$ for $\alpha=0.8$, $b=1$, and $k=1$ with different dimensions in FIG. 8. It can be observed that $M_{min}<2$ with $n<6$ at $Q=1$; however, the minimum inversion mass $2<M_{min}<2.5$ for $n=6$. Therefore, the inversion curve in FIG. 6(l) does not intersect with the constant enthalpy curve for $M=2$. The absence of an inversion point in the isenthalpic curves implies the absence of interconversion between the cooling and heating regions, indicating that the BH is uniformly heated. This suggests that the thermodynamics of black holes are affected by the dilaton parameter $\alpha$.  From FIG. 9, we can observe the influence of the horizon radius and pressure on the mass of the BH. As we are only investigating J-T expansion for the positive event horizon, the isenthalpic curves are real.

\begin{figure}[htbp]
	\centering
	\subfigure[$n=3$,$\alpha=0$]{
    \begin{minipage}[t]{0.32\linewidth}
		\centering
		\includegraphics[width=2in,height=1.2in]{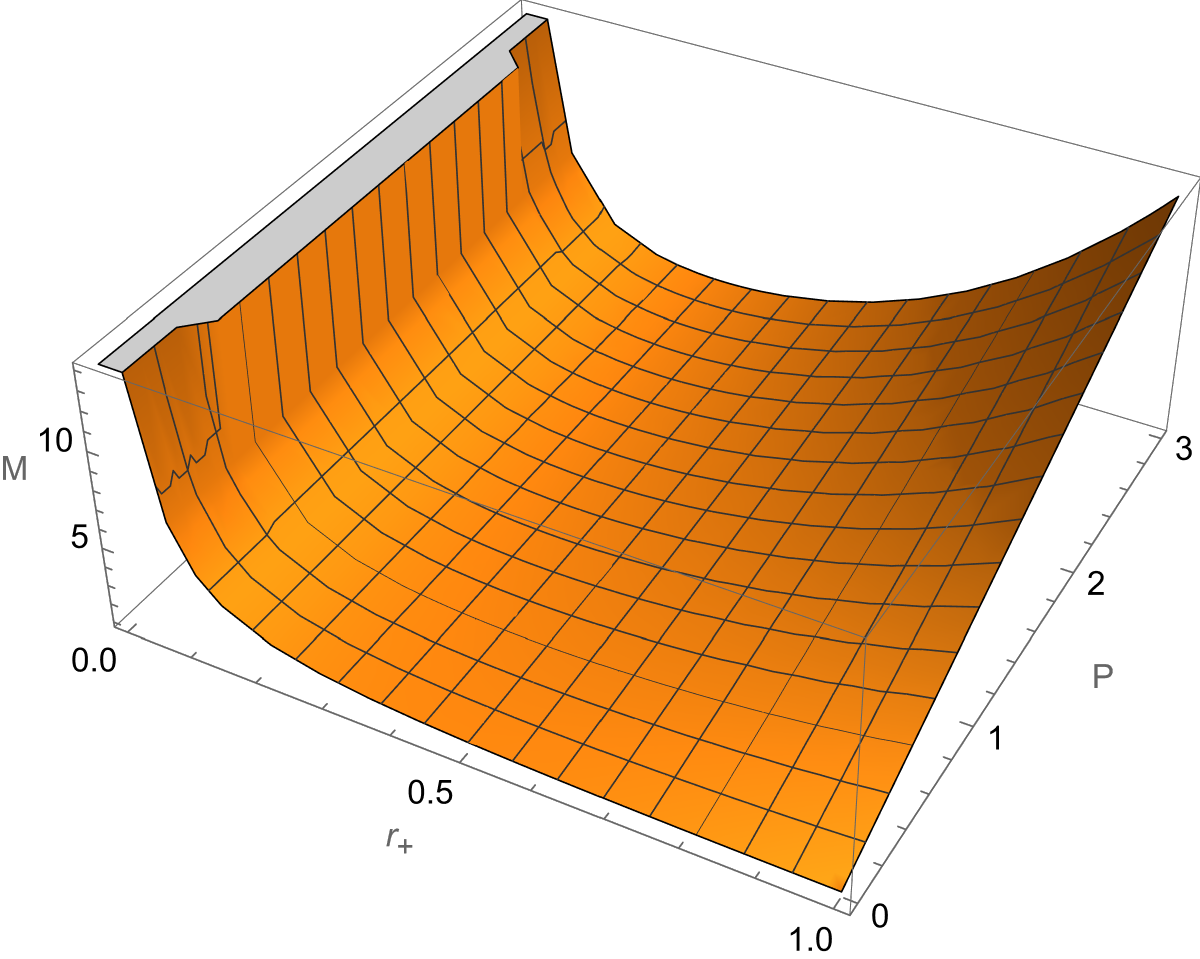}
		\end{minipage}%
            }%
    \subfigure[$n=3$,$\alpha=0.4$]{
    \begin{minipage}[t]{0.32\linewidth}
		\centering
		\includegraphics[width=2in,height=1.2in]{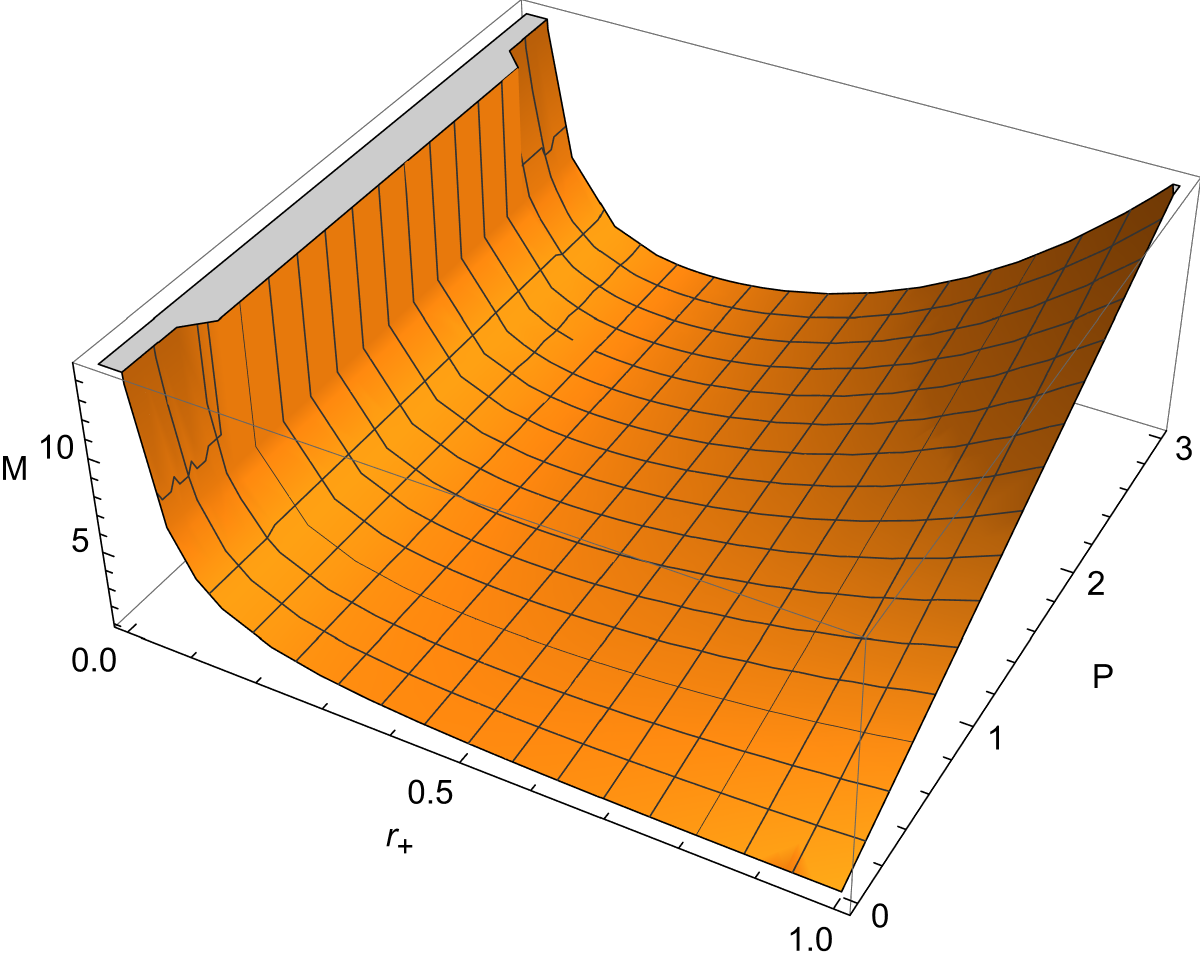}
		\end{minipage}%
            }%
      \subfigure[$n=3$,$\alpha=0.8$]{
    \begin{minipage}[t]{0.32\linewidth}
		\centering
		\includegraphics[width=2in,height=1.2in]{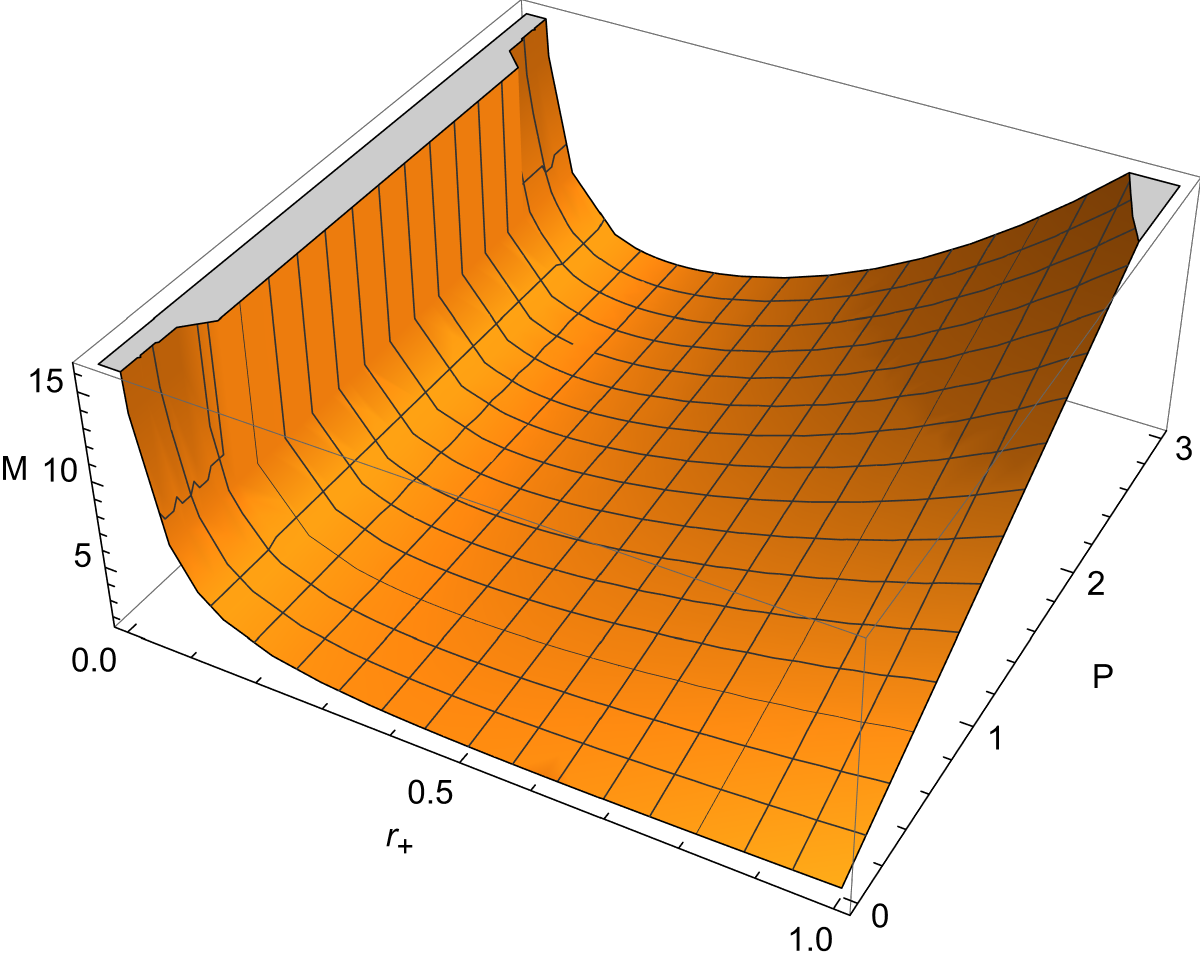}
		\end{minipage}%
            }%

            \subfigure[$n=4$,$\alpha=0$]{
    \begin{minipage}[t]{0.32\linewidth}
		\centering
		\includegraphics[width=2in,height=1.2in]{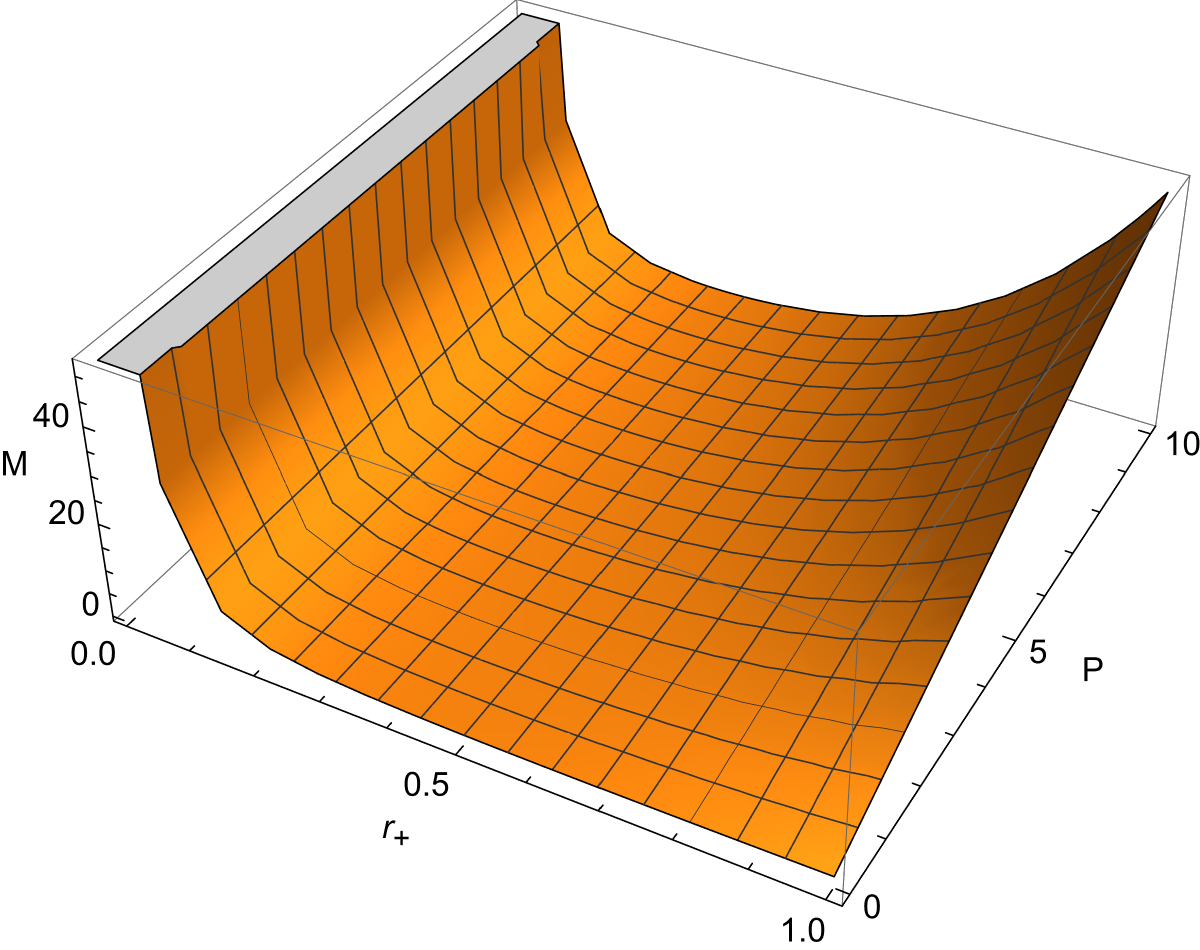}
		\end{minipage}%
            }%
    \subfigure[$n=4$,$\alpha=0.4$]{
    \begin{minipage}[t]{0.32\linewidth}
		\centering
		\includegraphics[width=2in,height=1.2in]{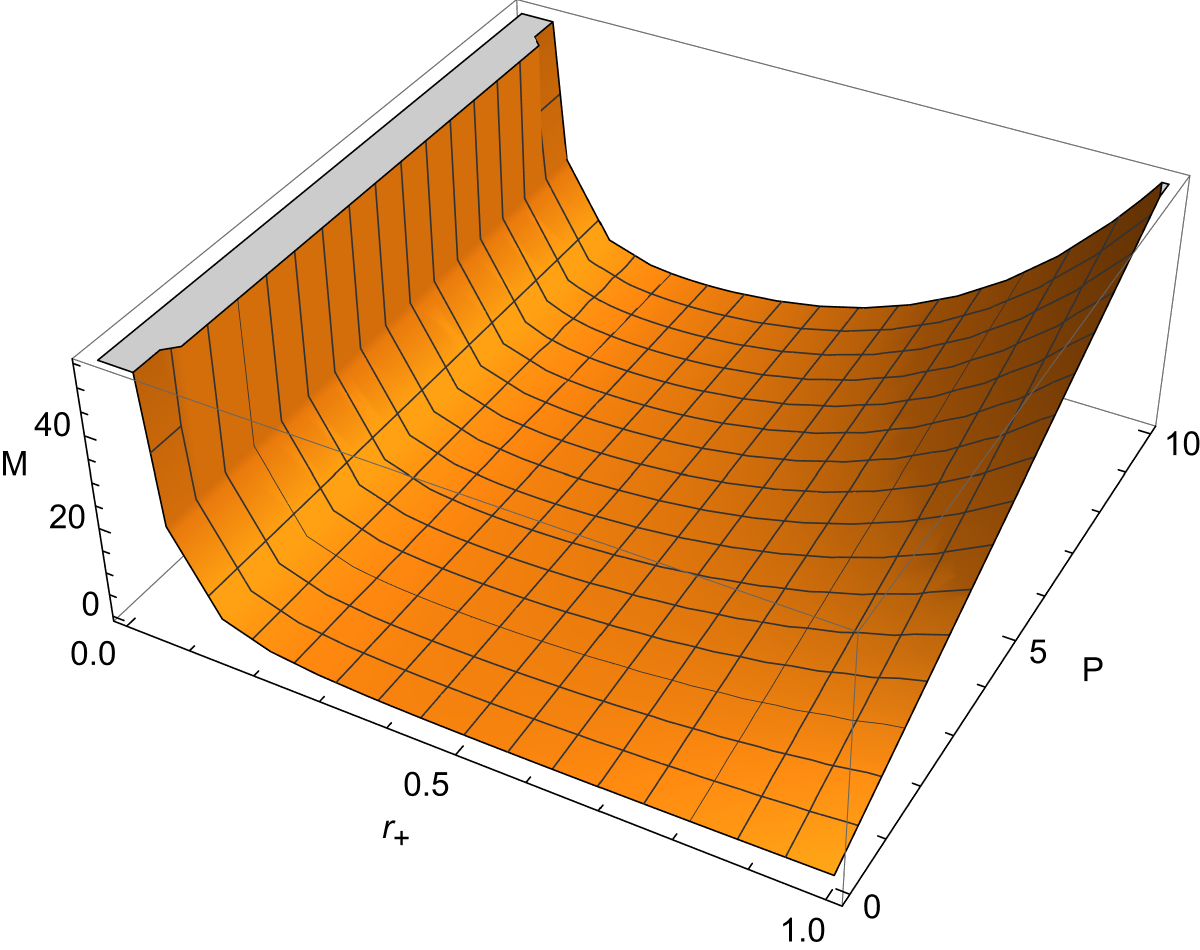}
		\end{minipage}%
            }%
      \subfigure[$n=4$,$\alpha=0.8$]{
    \begin{minipage}[t]{0.32\linewidth}
		\centering
		\includegraphics[width=2in,height=1.2in]{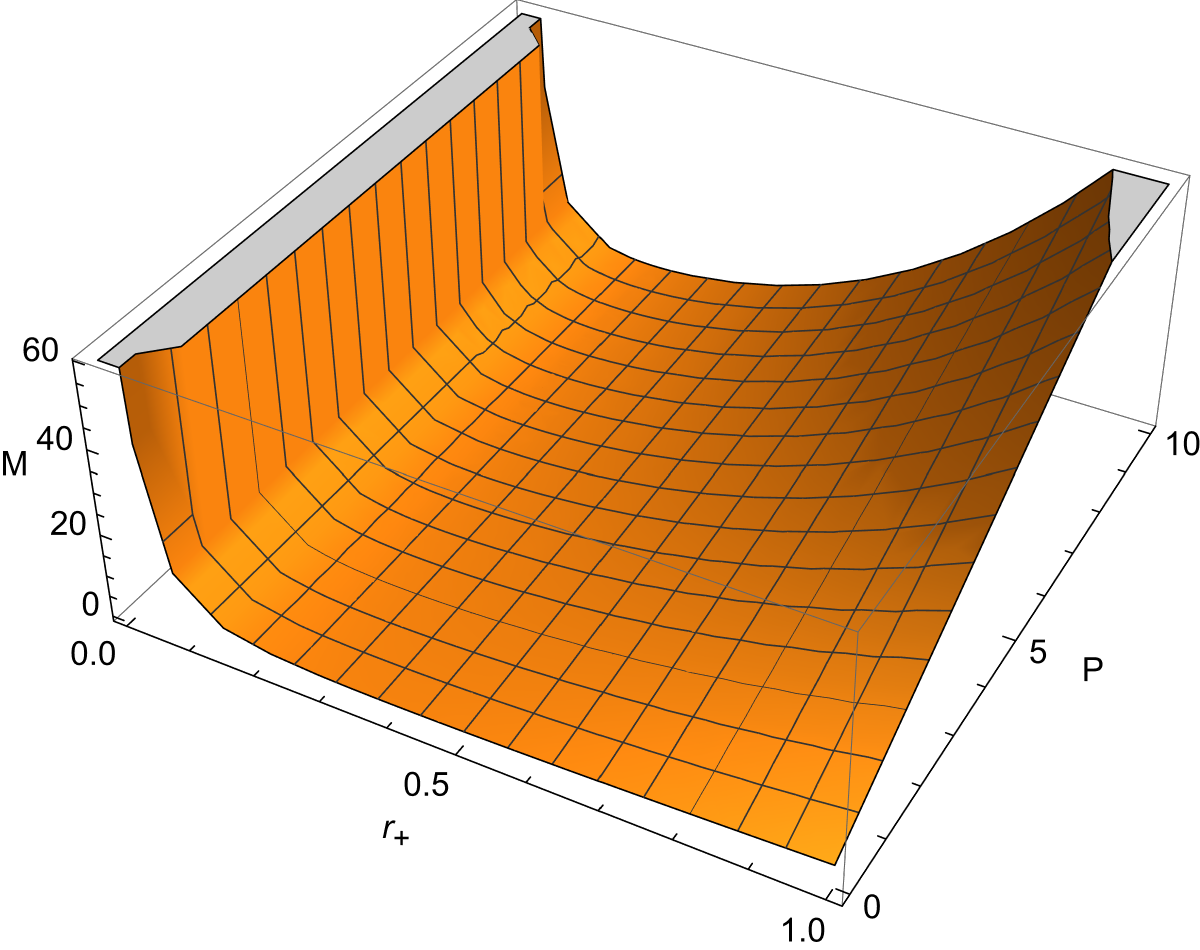}
		\end{minipage}%
            }%

            \subfigure[$n=5$,$\alpha=0$]{
    \begin{minipage}[t]{0.32\linewidth}
		\centering
		\includegraphics[width=2in,height=1.2in]{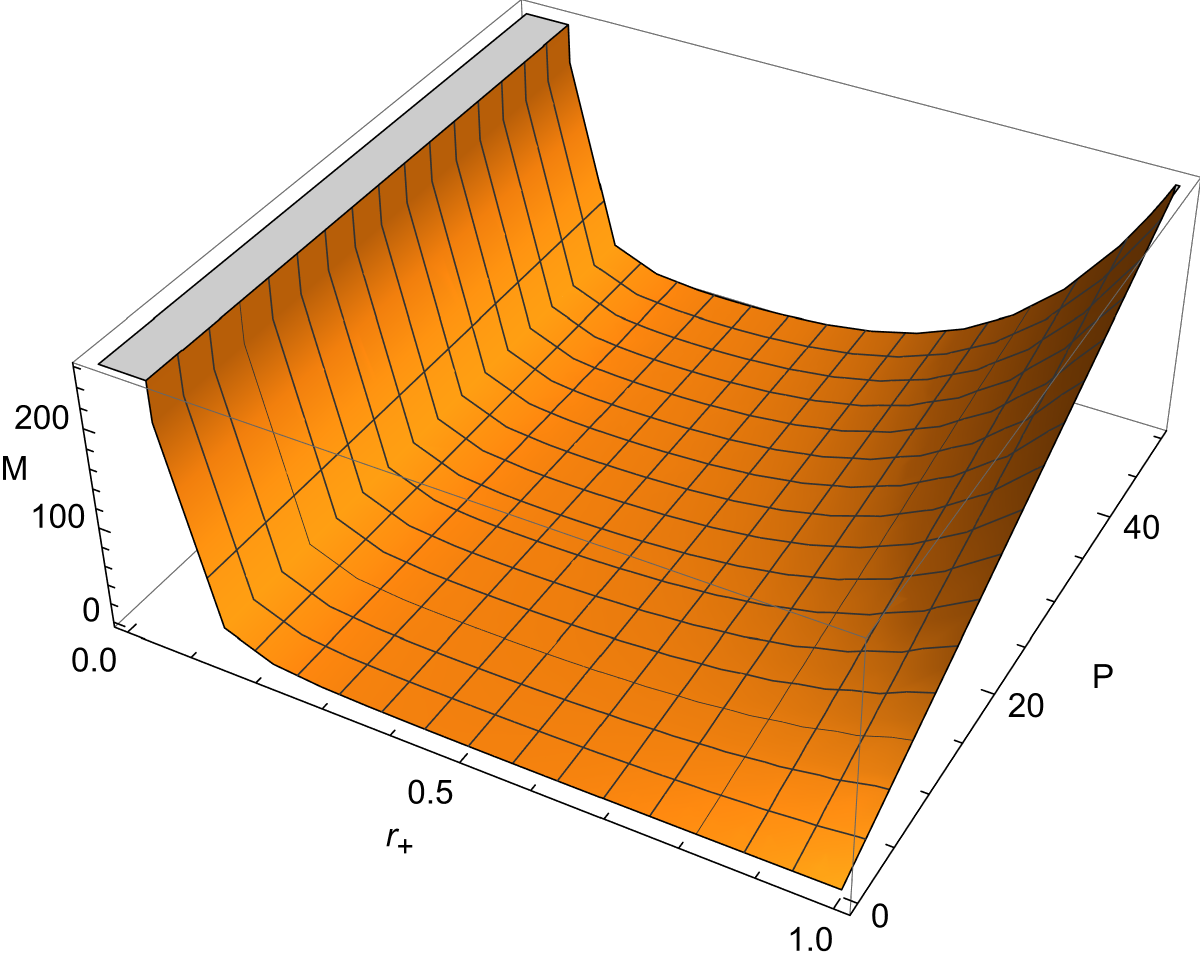}
		\end{minipage}%
            }%
    \subfigure[$n=5$,$\alpha=0.4$]{
    \begin{minipage}[t]{0.32\linewidth}
		\centering
		\includegraphics[width=2in,height=1.2in]{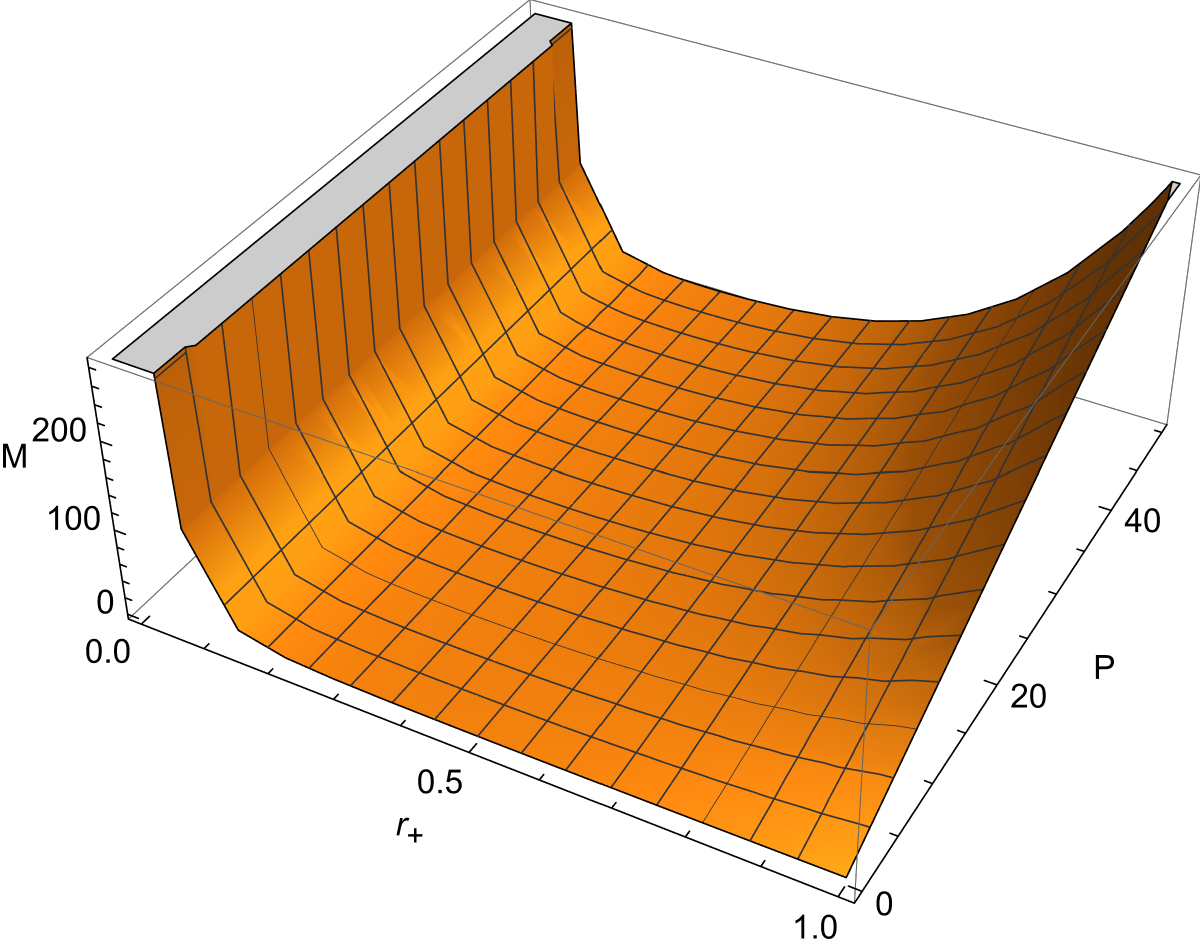}
		\end{minipage}%
            }%
      \subfigure[$n=5$,$\alpha=0.8$]{
    \begin{minipage}[t]{0.32\linewidth}
		\centering
		\includegraphics[width=2in,height=1.2in]{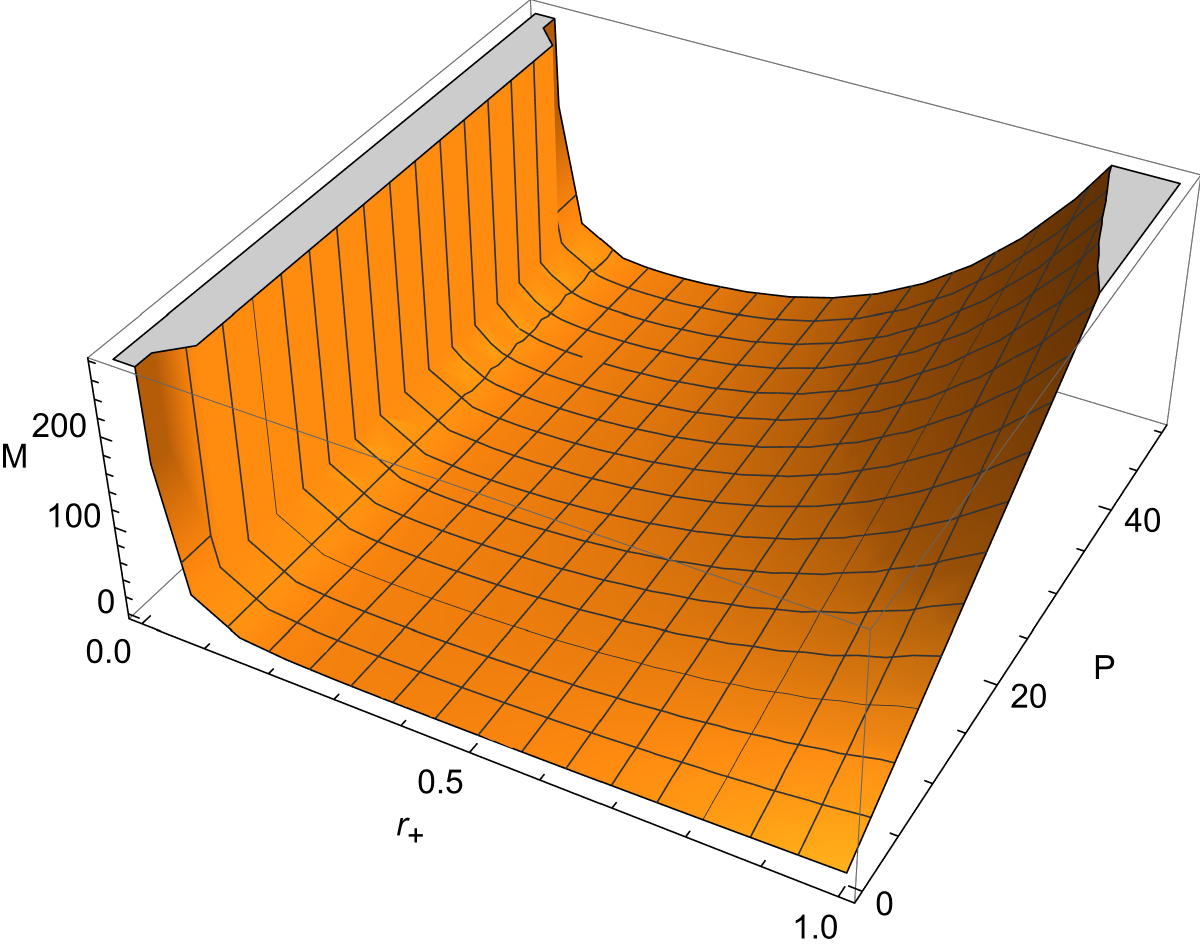}
		\end{minipage}%
            }%

         \subfigure[$n=6$,$\alpha=0$]{
    \begin{minipage}[t]{0.32\linewidth}
		\centering
		\includegraphics[width=2in,height=1.2in]{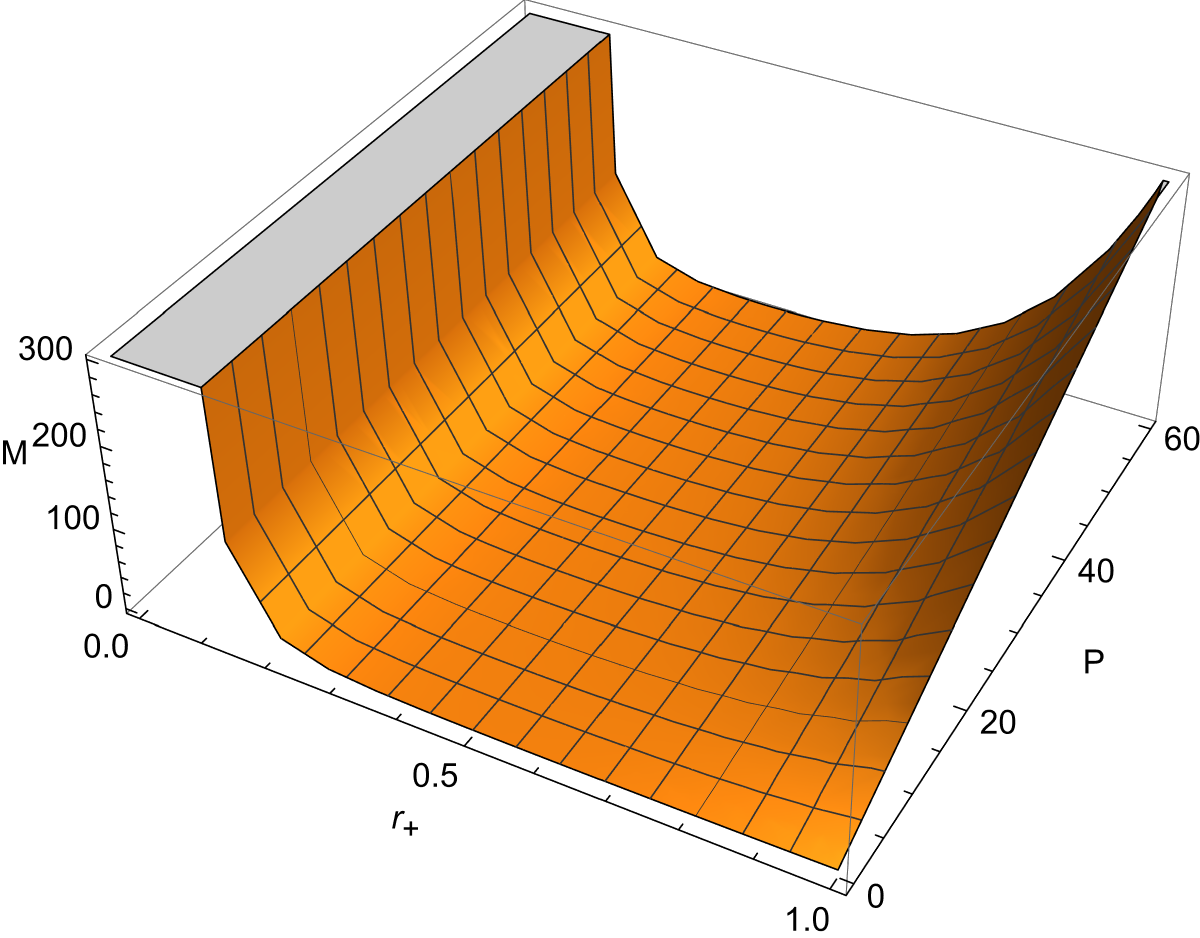}
		\end{minipage}%
            }%
    \subfigure[$n=6$,$\alpha=0.4$]{
    \begin{minipage}[t]{0.32\linewidth}
		\centering
		\includegraphics[width=2in,height=1.2in]{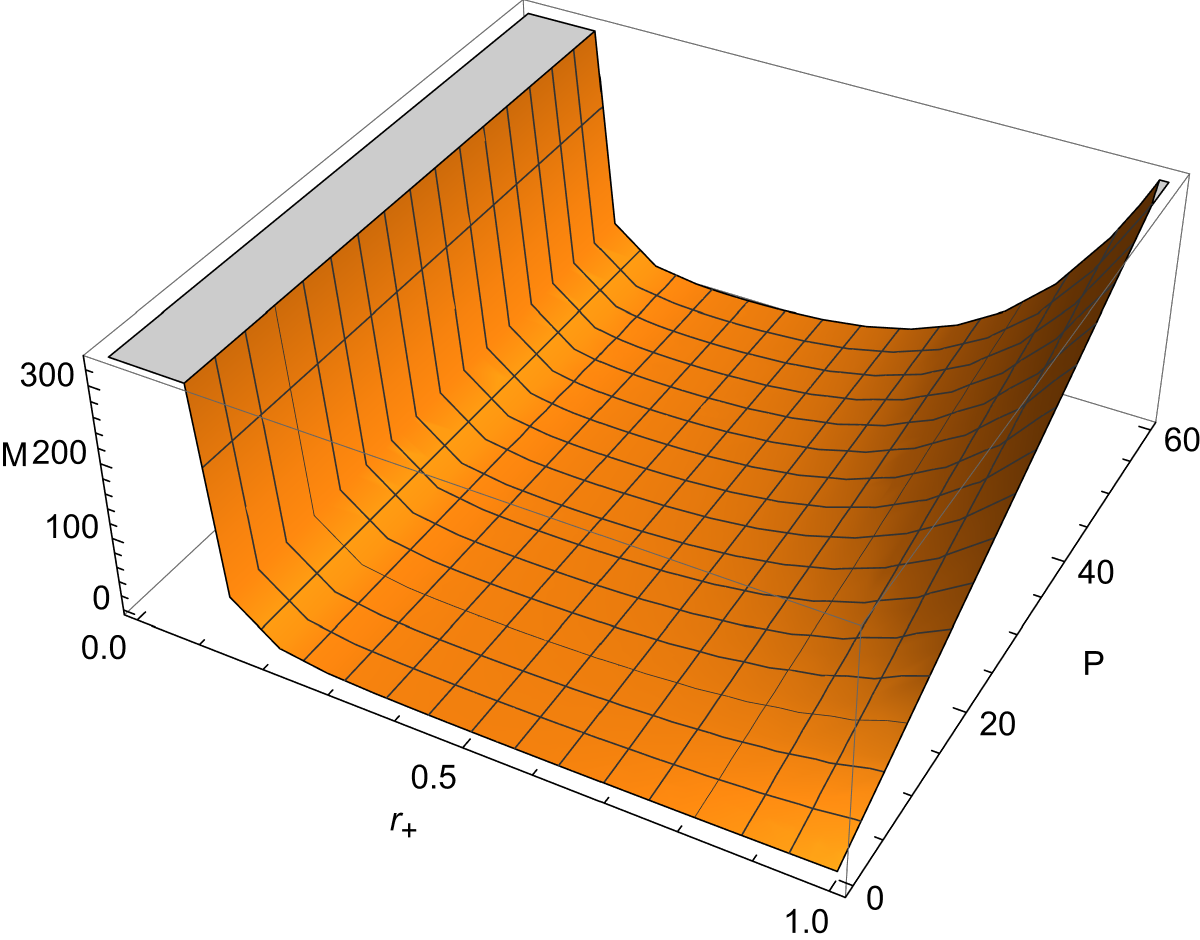}
		\end{minipage}%
            }%
      \subfigure[$n=6$,$\alpha=0.8$]{
    \begin{minipage}[t]{0.32\linewidth}
		\centering
		\includegraphics[width=2in,height=1.2in]{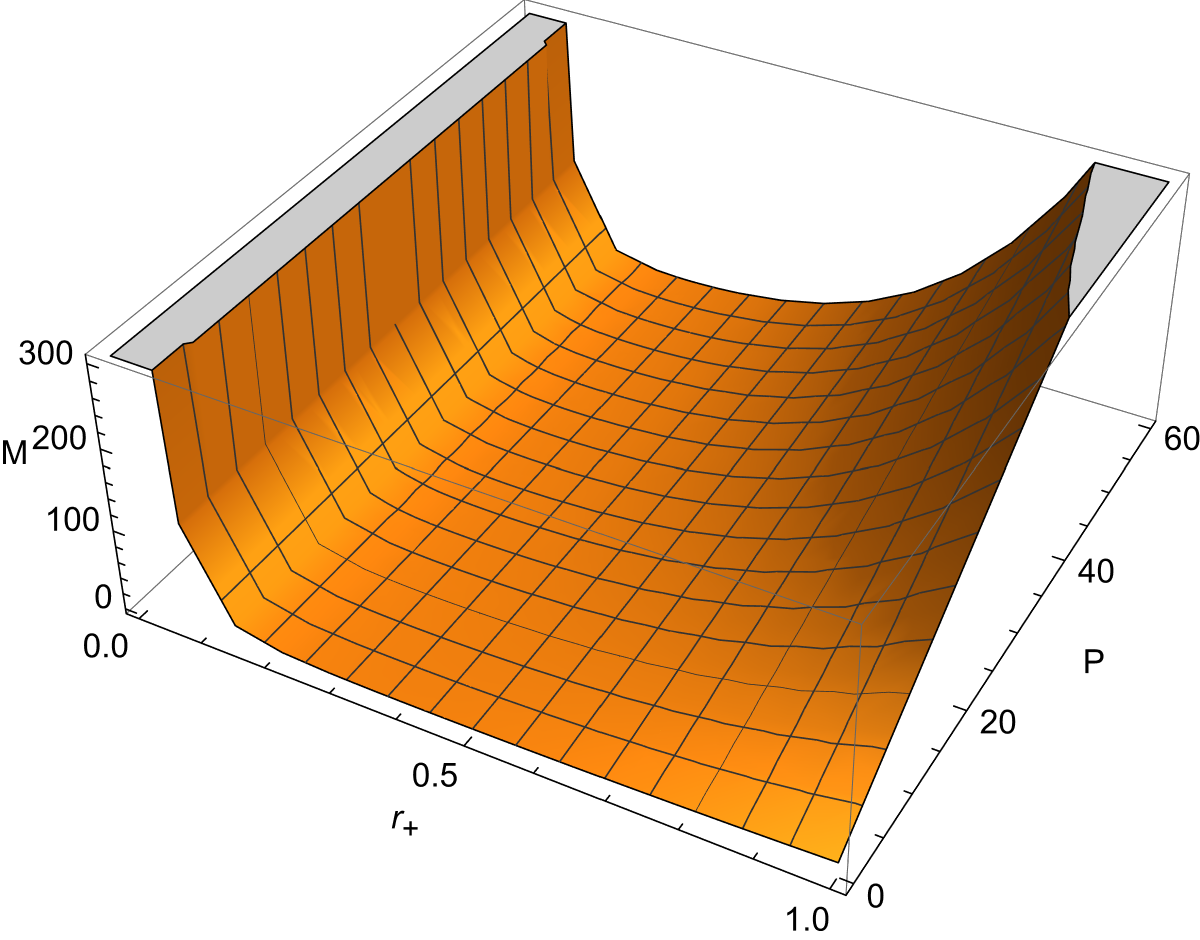}
		\end{minipage}%
            }%
     \centering
     \caption{Masses of charged dilatonic black holes versus the event horizon and pressure for $Q=1$, $b=1$, and $k=1$.}
\end{figure}
\section{CONCLUSIONS} \label{sec4}
In summary, J-T expansion describes an irreversible adiabatic expansion wherein the actual gas moves from a high pressure region to a low pressure region through a porous plug.  In this study, we primarily investigated the J-T expansion of $(n+1)$-dimensional charged dilatonic black holes in the presence of a potential $\mathcal{V}(\varphi)$. Consequently, we unveiled the effects of the dimension $n$ of space-time and dilaton field $\alpha$ on J-T expansion from various perspectives. The main results and findings are as follows.

First, we introduced the thermodynamic properties of charged dilatonic black holes in an extended phase space, along with the first law of thermodynamics and the Smarr relation. Next, we studied the J-T coefficient in detail, which is an important quantity in J-T expansion. The positive and negative J-T coefficients could be used to determine whether the gas was cooled or heated after an adiabatic expansion. We observed that the J-T coefficient had a critical point at a constant BH mass, and it may be interpreted as the divergent point of $\mu$. The corresponding horizon radius of the divergent point increased as the dimension $n$ increased but decreased as $\alpha$ increased. The same behavior could be observed at the zero point of the Hawking temperature $T$. In addition, we explored the possible influence of the heat capacity on J-T expansion and noted an interesting observation indicating that the negative heat capacity led to a cooling process.

Second, we analyzed the influence of BH characteristics on the inversion curve; these included the dimension($n$), electric charge(Q) and dilaton field($\alpha$). Our results revealed that in contrast to the effect of the dimension $n$, $T_{i}$ tended to decrease as $Q$ increased at low pressures, whereas the opposite result was obtained at high pressures. As increase in $n$ or $\alpha$ could lead to changes in the pressure cut-off points and the minimum inversion temperature $T_{min}$. Our analysis of inversion curves in the T-P plane revealed that the influence of the dilaton parameter $\alpha$ on a BH may be more evident in higher dimensional space-time. We deduced that these changes depended upon the characteristics of the BH.

Third, we numerically analyzed the ratio between the minimum inversion temperature $T_{min}$ and the critical temperature $T_{c}$, and the ratio was found to be independent of the charge of the BH but dependent on $\alpha$ and $n$. In addition, the ratio increased if we increased the dilaton parameter $\alpha$.  It was found that for $\alpha=0$ and $n=3$, the ratio was restored to 1/2 \cite{J1}. The value of the dilaton parameter altered the effect of the dimension on the ratio: for small values of $\alpha$, the ratio decreased with an increase in the dimension, but as $\alpha$ approached 1, the ratio was always greater than 1/2 and increased with an increase in the dimension.

Finally, because the enthalpy is known to remain constant during J-T expansion, we investigated the isenthalpic curve by fixing $Q$. A number of figures were plotted in the $T-P$ plane, indicating that the isenthalpic curves crossed the inversion curves. In these figures, we could observe that the inversion curves coincided with the extreme point of a specific isenthalpic curve, and the cooling-heating regions could be identified. In other words, the boundary between the heating and cooling regions of the BH depends on the inversion curve. We also discovered that both the maximum expansion point of the isenthalpic curves and the cooling-heating critical point would change in the $T-P$ plane. To further explain why the inversion curve in FIG. 6(l) did not intersect with the constant enthalpy curve for $M=2$, we plotted the behavior of the minimum inversion mass $M_{min}$. In addition, the behavior of the mass versus the event horizon and pressure was visualized. In summary, the dimension($n$) and dilaton field parameter($\alpha$) play an important role in J-T expansion.
\section*{Acknowledgments}
 The authors would like to thank the kind refere for their constructive suggestions on improving our paper. This work is supported by the National Natural Science Foundation of China (11465006 and 11565009) and the Doctoral Foundation of Zunyi Normal University of China (BS [2022] 07, QJJ-[2022]-314).





\begin{thebibliography}{0}
\expandafter\ifx\csname natexlab\endcsname\relax\def\natexlab#1{#1}\fi
\expandafter\ifx\csname bibnamefont\endcsname\relax
  \def\bibnamefont#1{#1}\fi
\expandafter\ifx\csname bibfnamefont\endcsname\relax
  \def\bibfnamefont#1{#1}\fi
\expandafter\ifx\csname citenamefont\endcsname\relax
  \def\citenamefont#1{#1}\fi
\expandafter\ifx\csname url\endcsname\relax
  \def\url#1{\texttt{#1}}\fi
\expandafter\ifx\csname urlprefix\endcsname\relax\def\urlprefix{URL }\fi
\providecommand{\bibinfo}[2]{#2}
\providecommand{\eprint}[2][]{\url{#2}}

\end{thebibliography}


\begin{thebibliography}{99}
\bibitem{P1}
J. D. Bekenstein, Phys. Rev. D 7 (1973) 2333.
\bibitem{P2}
J. M. Bardeen, B. Carter, and S. W. Hawking, Commun. Math. Phys. 31 (1973) 161.
\bibitem{P3}
S. W. Hawking, Commun. Math. Phys. 43 (1975) 199.

\bibitem{Q1}
R. M. Wald, Living. Rev. Relativ. 4 (2001) 6.
\bibitem{Q2}
D. N. Page, New. J. Phys. 7 (2005) 203.
\bibitem{Q3}
S. W. Hawking and D. N. Page, Commun. Math. Phys. 87 (1983) 577.
\bibitem{Q4}
J. D. Bekenstein, Phys. Rev. D 9 (1974) 3292.

\bibitem{ch1}
W. S. Chung and H. Hassanabadi, Phys. Lett. B 793 (2019) 451.
\bibitem{ch2}
H. Hassanabadi, E. Maghsoodi and W. S. Chung, Eur. Phys. J. C 79 (2019) 358.
\bibitem{ch3}
H. Chen, B. C. L\"utf\"uoglu, H. Hassanabadi and Z. W. Long, Phys. Lett. B 827 (2022) 136994.
\bibitem{ch4}
Y.~Kumaran and A.~\"Ovg\"un,   Chin. Phys. C 44 (2020) 025101.


\bibitem{E1}
G. t Hooft, Nucl. Phys. B 256 (1985) 727.
\bibitem{E2}
J. L. Cardy, Nucl. Phys. B 270 (1986) 186.

\bibitem{PT1}
S. W. Wei and Y. X. Liu, Phys. Rev. D 90 (2014) 044057.
\bibitem{PT2}
P. Chen and R. J. Adler, Nucl. Phys. B 124 (2003) 103.
\bibitem{PT3}
D. W. Yan, Z. R. Huang, N. Li, Chin. Phys. C 45 (2021) 015104.

\bibitem{R1}
J. Y. Zhang and J. H. Fan, Phys. Lett. B 648 (2007) 133.
\bibitem{R2}
X. X. Zeng and S. Z. Yang, Gen. Rel. Grav. 40 (2008) 2107.
\bibitem{R3}
X. X. Zeng, J. S. Hou and S. Z. Yang, Pramana 70 (2008) 409.

\bibitem{C1}
S. H. Hendi, Z. Armanfard, Gen. Relat. Gravit. 47 (2015) 125.
\bibitem{C2}
S. H. Hendi, K. Jafarzade, Phys. Rev. D 103 (2021) 104011.
\bibitem{C3}
E. Spallucci, A. Smailagic, Phys. Lett. B 723 (2013) 436.
\bibitem{C4}
S. W. Wei and Y. X. Liu, Phys. Rev. D 87 (2013) 044014.
\bibitem{C5}
D. C. Zou, S. J. Zhang, B. Wang, Phys. Rev. D  89 (2014) 044002.
\bibitem{C6}
Y. B. Ma, R. Zhao, S. Cao, Eur. Phys. J. C 76 (2016) 1.
\bibitem{C7}
A. Dehyadegari, A.Sheykhi, A. Montakhab, Phys. Lett. B 768 (2017) 235.

\bibitem{A1} D. Kastor, S. Ray, J. Traschen, Class. Quant. Grav. 26 (2009) 195011.
\bibitem{A2} D. Kastor, S. Ray, J. Traschen, Class. Quant. Grav. 27 (2010) 235014.
\bibitem{A3} D. Kastor, S. Ray, J. Traschen, Class. Quant. Grav. 28 (2011) 195022.
\bibitem{A4} D. Kastor, S. Ray, J. Traschen, Class. Quant. Grav. 36 (2018) 024002.

\bibitem{M1}
S. H. Hendi and M. Vahidinia, Phys. Rev. D  88 (2013) 084045.
\bibitem{M2}
A. M. Frassino, D. Kubiz\u{n}\'ak, R. B. Mann and F. Simovic, JHEP. 80 (2014) 2014.
\bibitem{F1}
D. Kubiznak and B. R. Mann, JHEP. 07 (2012) 033.
\bibitem{F2}
A. Belhaj, M. Chabab, H. El Moumni, and M. B. Sedra, Chin. Phys. Lett. 29 (2012) 100401.
\bibitem{F3}
S. H. Hendi, B. Eslam Panah, and S. Panahiyan, Phys. Lett. B 769 (2017) 191.

\bibitem{ADS}
J. Maldacena,  Int. J. Theor. Phys. 38 (1999) 1113.

\bibitem{BH1}
A. Belhaj, M. Chabab, H. El Moumni, L. Medari, and M. B. Sedra, Chin. Phys. Lett. 30 (2013) 090402.
\bibitem{BH2}
A. Belhaj, M. Chabab, H. El Moumni, K. Masmar, and M. B. Sedra, Int. J. Geom. Meth. Mod. Phys. 12 (2014) 1550017.

\bibitem{MT1}
A. Belhaj, M. Chabab, H. El Moumni, K. Masmar, and M. B. Sedra, Eur. Phys. J. C 76 (2016) 73.
\bibitem{MT2}
M. Chabab, H. El Moumni, and K. Masmar, Eur. Phys. J. C 76 (2016) 304.

\bibitem{QU1}
M. Chabab, H. El Moumni, S. Iraoui, and K. Masmar, Eur. Phys. J. C 76 (2016) 676.
\bibitem{QU2}
M. Chabab, H. El Moumni, S. Iraoui, and K. Masmar, Astrophys. Space Sci. 362 (2017) 192.

\bibitem{HE1}
A. Belhaj, M. Chabab, H. El Moumni, K. Masmar, M. B. Sedra, and A. Segui, JHEP. 05 (2015) 149.

\bibitem{ST1}
M. Chabab, H. El Moumni, S. Iraoui, K. Masmar, and S. Zhizeh, Phys. Lett. B 781 (2018) 316.

\bibitem{J1}
\"O. \"Okc\"u and E. Aydiner, Eur. Phys. J. C 77 (2017) 24.
\bibitem{J2}
\"O. \"Okc\"u and E. Aydiner, Eur. Phys. J. C 78 (2018) 123.
\bibitem{J3}
J. X. Mo, G. Q. Li, S.Q. Lan, X.B. Xu, Phys. Rev. D 98 (2018) 124032.
\bibitem{J4}
J. X. Mo, G. Q. Li, Class. Quantum. Grav. 37 (2020) 045009.
\bibitem{J5}
S. Q. Lan, Phys. Rev. D 98 (2018) 084014.
\bibitem{J6}
J. T. Xing, Y. Meng, X. M. Kuang, Phys. Lett. B 820 (2021) 136604.
\bibitem{J7}
J. Liang, B. Mu, P. Wang, Phys. Rev. D 104 (2021) 124003.
\bibitem{J8}
A. Ditta, X. Tiecheng, G. Mustafa, et al. Eur. Phys. J. C 82 (2022) 1.

\bibitem{RE1} I. Z. Fisher, Zh. Eksp. Teor. Fiz. 18 (1948) 636. [Sov. Phys. JETP 18 (1948) 636].
\bibitem{RE2} M. B. Green, J. H. Schwartz, E. Witten, Superstring Theory, CUP (1987)
\bibitem{RE3} B. Harms and Y. Leblanc, Phys. Rev. D 46 (1992) 2334.
\bibitem{RE4} K. C. K. Chan, J. H. Horne and R. B. Mann, Nucl. Phys. B 447 (1995) 441.
\bibitem{RE5} A. Sheykhi, Phys. Rev. D 76 (2007) 124025.
\bibitem{RE6} S. H. Hendi, A. Sheykhi, S. Panahiyan, et al, Phys. Rev. D  92 (2015) 064028.
\bibitem{RE7} A. Dehyadegari, A. Sheykhi, A. Montakhab, Phys. Rev. D  96 (2017) 084012.
\bibitem{RE8} R. Zhao, H. H. Zhao, M. S. Ma and L. C. Zhang, Eur. Phys. J. C 73 (2013) 2645.
\bibitem{RE9} J. X. Mo, S. Q. Lan, Chin. Phys. C 45 (2021) 105106.
\bibitem{RE10} M. H. Dehghani, S. Kamrani, A. Sheykhi, Phys. Rev. D 90 (2014) 104020.
\bibitem{RE12} M. U. Shahzad, L. Nosheen, Eur. Phys. J. C 82 (2022) 1.
\bibitem{RE13} S. Chaudhary, A. Jawad, M. Yasir, Phys. Rev. D 105 (2022) 024032.


\bibitem{Smarr1}
L. Smarr, Phys. Rev. Lett. 30 (1973) 71.



\end{thebibliography}
\end{document}